\documentclass[letterpaper,titlepage,11pt]{article}

\pdfoutput=1

\usepackage{amssymb,amsmath,amsfonts, graphics}
\usepackage{bm}
\usepackage{epsfig}
\usepackage{ccaption}
\usepackage{graphicx}

\usepackage[
      colorlinks=true,
      linkcolor=blue,
      urlcolor=blue,
      filecolor=black,
      citecolor=red,
      pdfstartview=FitV,
      pdftitle={},
        pdfauthor={Roberto Emparan, Raimon Luna, Keisuke Izumi, Ryotaku Suzuki, Kentaro Tanabe},
        pdfsubject={},
        pdfkeywords={},
        pdfpagemode=None,
        bookmarksopen=true
      ]{hyperref}

\def\clock{{\count0=\time
           \divide\count0 60
           \ifnum\count0<10 0\fi\the\count0
           \multiply\count0 -60 \advance\count0 \time
           :\ifnum\count0<10 0\fi \the\count0
         }}
\newcommand{\timestamp}{{\small\vbox{\hbox{\tt\jobname.tex}
\hbox{\the\day/\the\month/\the\year, \clock}}}}

\newcommand{\ie}{{\it i.e.,\,}}
\newcommand{\eg}{{\it e.g.,\,}}

\newcommand{\lp}{\left(}
\newcommand{\rp}{\right)}

\newcommand{\mc}[1]{\mathcal{#1}}

\newcommand{\beq}{\begin{equation}}
\newcommand{\eeq}{\end{equation}}
\newcommand{\bea}{\begin{eqnarray}}
\newcommand{\eea}{\end{eqnarray}}
\newcommand{\beqa}{\begin{eqnarray}}
\newcommand{\eeqa}{\end{eqnarray}}

\newcommand{\sR}{\mathsf{R}}

\newcommand{\cR}{\mathcal{R}}

\newcommand{\bsigma}{\boldsymbol{\sigma}}
\newcommand{\brho}{\boldsymbol{\rho}}
\newcommand{\bT}{\boldsymbol{T}}
\newcommand{\bq}{\boldsymbol{q}}
\newcommand{\bPhi}{\boldsymbol{\Phi}}
\newcommand{\bfeta}{\boldsymbol{\eta}}
\newcommand{\bkappa}{\boldsymbol{\kappa}}
\newcommand{\bomega}{\boldsymbol{\omega}}
\newcommand{\bk}{\boldsymbol{k}}
\newcommand{\bv}{\boldsymbol{v}}
\newcommand{\bP}{\boldsymbol{P}}
\newcommand{\bs}{\boldsymbol{s}}
\newcommand{\bmu}{\boldsymbol{\mu}}

\newcommand{\fr}[1]{\frac{1}{#1}}

\newcommand{\ord}[1]{{\mathcal O}(#1)}

\newcommand{\cE}{{\mathcal E}}

\newcommand{\cJ}{{\mathcal J}}
\newcommand{\cL}{{\mathcal L}}

\newcommand{\nonum}{\nonumber\\ }

\numberwithin{equation}{section}

\begin{document}

\begin{titlepage}
\rightline{KEK-TH-1885, OCU-PHYS-438, AP-GR-130} 
\leftline{}
\vskip 1.5cm
\centerline{\LARGE \bf Hydro-elastic Complementarity}
\medskip
\centerline{\LARGE \bf in Black Branes at large $D$} 
\vskip 1.cm
\centerline{\bf Roberto Emparan$^{a,b}$, Keisuke Izumi$^{b}$, Raimon Luna$^{b}$,} 
\centerline{\bf Ryotaku Suzuki$^{c}$, Kentaro Tanabe$^{d}$}
\vskip 0.5cm
\centerline{\sl $^{a}$ICREA, Passeig Llu\'{\i}s Companys 23, E-08010 Barcelona, Spain}
\smallskip
\centerline{\sl $^{b}$Departament de F{\'\i}sica Fonamental, Institut de
Ci\`encies del Cosmos,}
\centerline{\sl  Universitat de
Barcelona, Mart\'{\i} i Franqu\`es 1, E-08028 Barcelona, Spain}
\smallskip
\centerline{\sl $^{c}$Department of Physics, Osaka City University, Osaka 558-8585, Japan}
\smallskip
\centerline{\sl $^{d}$Theory Center, Institute of Particles and Nuclear Studies, KEK,}
\centerline{\sl  Tsukuba, Ibaraki, 305-0801, Japan}
\smallskip
\vskip 0.5cm
\centerline{\small\tt emparan@ub.edu,\, keisuke.aset@gmail.com,\, raimonluna@gmail.com,} \centerline{\small\tt ryotaku@sci.osaka-cu.ac.jp,\, ktanabe@post.kek.jp}

\vskip 1.0cm
\centerline{\bf Abstract} \vskip 0.2cm \noindent

We obtain the effective theory for the non-linear dynamics of black branes---both neutral and charged, in asymptotically flat or Anti-deSitter spacetimes---to leading order in the inverse-dimensional expansion. We find that black branes evolve as viscous fluids, but when they settle down they are more naturally viewed as solutions of an elastic soap-bubble theory. The two views are complementary: the same variable is regarded in one case as the energy density of the fluid, in the other as the deformation of the elastic membrane. The large-$D$ theory captures finite-wavelength phenomena beyond the conventional reach of hydrodynamics. For asymptotically flat charged black branes (either Reissner-Nordstrom or $p$-brane-charged black branes) it yields the non-linear evolution of the Gregory-Laflamme instability at large $D$ and its endpoint at stable non-uniform black branes. For Reissner-Nordstrom AdS black branes we find that sound perturbations do not propagate (have purely imaginary frequency) when their wavelength is below a certain charge-dependent value. We also study the polarization of black branes induced by an external electric field.

\end{titlepage}
\pagestyle{empty}
\small
\normalsize
\newpage
\pagestyle{plain}
\setcounter{page}{1}



\section{Introduction and summary}

The limit of large number of dimensions $D$ concentrates the gravitational field of a black hole within a short distance $\sim 1/D$ of its horizon, leaving an undistorted background outside it \cite{Asnin:2007rw,Emparan:2013moa,Emparan:2013xia}. It is then natural to expect that the black hole and its field can be replaced by a thin, effective  membrane-like object in the background geometry. Equations for this effective membrane have been recently derived in \cite{Emparan:2015hwa,Bhattacharyya:2015dva,Suzuki:2015iha,Suzuki:2015axa,Emparan:2015gva,Tanabe:2015hda,Bhattacharyya:2015fdk}, in different formulations and in regimes that overlap but do not entirely coincide. In particular, ref.~\cite{Emparan:2015hwa} obtained fully covariant equations for static black holes,  both in vacuum ($R_{\mu\nu}=0$) and in (Anti-)deSitter ($R_{\mu\nu}=\pm\Lambda g_{\mu\nu}$), including leading order and next-to leading order terms in the $1/D$ expansion. This allows to consider fluctuations on horizon scales of order one, but also of order $1/\sqrt{D}$, as is appropriate, as we explain below, for black branes. These simple covariant equations were extended in ref.~\cite{Suzuki:2015iha} to stationary black holes, but now restricted to vacuum  and to leading order in the expansion. The first formulation to include time dependence was achieved in ref.~\cite{Bhattacharyya:2015dva}, working to leading order and for vacuum black holes. It can describe horizon shapes that vary on length and time scales of order one, and velocities along the horizon that are also of order one --- hence not yet capable of capturing black brane dynamics. As it happens, in order to obtain the time-dependent effective theory for the latter, rather than calculating the next-order corrections to the equations of \cite{Bhattacharyya:2015dva}, it is more efficient to solve for a black brane ansatz  which readily yields a set of simple effective equations (for vacuum or Anti-deSitter) that can be easily studied and solved \cite{Emparan:2015gva}. Similar specific ansatze have permitted to obtain new results for black strings and black rings \cite{Suzuki:2015axa,Tanabe:2015hda}. In a sense, the existence of all these formulations reveals a trade-off between, on the one hand, the goal of a formulation with the highest generality, and on the other hand, the desirability of simple equations for specific systems, which are easy to derive and to use for obtaining new results. Hopefully, further advances will combine the best of all these approaches.

In all cases, the theory is naturally formulated using geometric variables that describe how the membrane is embedded in the background, plus velocity fields for motion (\eg\ rotation) along the spatial directions of the horizon. Charge on the black hole brings in an additional effective field for the charge distribution on the membrane \cite{Bhattacharyya:2015fdk}.\footnote{Refs.~\cite{Emparan:2013oza,Tanabe:2015isb,Guo:2015swu,Andrade:2015hpa} study other aspects of charged black holes in the $1/D$ expansion.}

A basic question is whether the equations of the effective theory can be understood in terms of familiar physics. 
Previous results suggest two different kinds of interpretation: 
\begin{itemize}
\item Refs.~\cite{Emparan:2015hwa,Suzuki:2015iha} find that stationary black holes at large $D$ are solutions of an \textit{elastic} theory: the effective membrane must satisfy the equation
\beq\label{soapeq}
K=2\gamma\kappa\,,
\eeq
where $K$ is the trace of the extrinsic curvature of the membrane, the factor
\beq
\gamma^{-1}=\sqrt{-g_{tt}(1-v^2)}
\eeq
accounts for the gravitational and Lorentz boost redshifts on the membrane, and the constant $\kappa$ is the surface gravity of the black hole. When $\gamma=1$ (\eg\ static black holes in a Minkowski background) this is the familiar Young-Laplace equation for soap bubbles (membrane interfaces). 

\item The effective theory of large-$D$ black branes must contain \textit{hydrodynamic} features, since it is the non-linear theory of the lowest-frequency quasinormal modes of the brane in the large-$D$ limit. These have been shown in \cite{Emparan:2014aba,Emparan:2015rva} to be precisely, and only, the modes that at long-wavelengths produce hydrodynamic behavior on the brane \cite{Kovtun:2003wp,Bhattacharyya:2008jc,Camps:2010br}. 
\end{itemize}
How, then, are these elastic and hydrodynamic viewpoints reconciled? Put pictorially: are ripples on a black brane like pressure waves on a fluid, or rather like wrinkles on a membrane?

In this article we reformulate and extend the results in \cite{Emparan:2015gva} to obtain the large-$D$ effective theory of black branes---now including charge, in asymptotically flat (AF) or Anti-deSitter (AdS) spacetimes---in a formulation that incorporates the two perspectives in a complementary way. That is, we can write the equations for the dynamical evolution of the black brane in hydrodynamic form, but for stationary configurations they are more naturally viewed as soap-bubble-type equations like \eqref{soapeq}. These two views (none of which were manifest in  \cite{Emparan:2015gva}) are complementary in that the same variable is interpreted, in one case as the energy density of the fluid, in the other case as the `height function' measuring the deformation of an elastic membrane.\footnote{Despite similarities in wording, this is very different from the blackfold effective fluid on an elastic membrane \cite{Emparan:2009at}, and also from the usual membrane paradigm \cite{Damour:1978cg}, which does not possess these elastic aspects. }
 Indeed this reflects a basic feature of black holes: the same variable that gives their mass also sets the horizon size.

In more detail, by solving the Einstein equations for black branes to leading order in the $1/D$ expansion, we obtain a hydrodynamic theory of a compressible, viscous fluid, with a conserved particle-number current (from the charge density on the brane). Even though the large-$D$ expansion and the hydrodynamic gradient expansion are different, the effective fluid stress tensor contains only a finite number of terms beyond the viscous stress, indeed just a single term with two derivatives of the mass density. Reinterpreting the mass density variable as the local radius of the horizon, this term completes the extrinsic curvature for a membrane embedded in a background and yields eq.~\eqref{soapeq} for stationary configurations.

We restrict our analysis to black brane configurations, with spatial horizon topology $\mathbb{R}^{p}\times S^{D-p-2}$ in AF and $\mathbb{R}^{D-2}$ in AdS. The fluctuations occur along the extended brane directions and have wavelengths $\sim 1/\sqrt{D}$ (which is the scaling of the sound speed on the brane). For AF branes the sphere is kept metrically round, and only its size varies. It might seem desirable to have a more covariant formulation that can account for fluctuations on scales $\sim 1/\sqrt{D}$ of the sphere, and more generally not only of black branes, but also of black holes with, \eg\ a topologically spherical horizon $S^{D-2}$. However, the analysis of linearized perturbations of spherical black holes reveals that quasinormal modes with partial wave numbers $\ell\sim \sqrt{D}$ do not decouple from the far-zone background, so they are not amenable to an effective theory \cite{Emparan:2014aba}. Therefore, the restricted covariance of our ansatz seems to be an inevitable feature of the effective dynamics for horizon wavelengths $\sim 1/\sqrt{D}$.

In addition to black branes with a charge coupled to a Maxwell-type $1$-form potential, often dubbed Reissner-Nordstrom (RN) black branes, we have also obtained the effective theory of black $p$-branes charged under a $(p+1)$-form potential. This extended $p$-brane charge cannot be redistributed along the worldvolume and therefore is not associated to any local degree of freedom for a particle-number density in the effective theory. 
Both types of charge modify the tension of the brane, and hence the elasticity of the effective membrane. They also change the pressure of the effective fluid, and can alter its stability, as indeed we find.  

Additionally, it is also easy to incorporate an external electric field parallel to the brane, and study the polarizing effect it has on the brane. Similar black hole polarization in global AdS has been recently studied in \cite{Costa:2015gol}.

The large-$D$ effective theory of black branes and the hydrodynamic fluid/gravity theories relate and compare to each other in several respects:
\begin{itemize}
\item The theories coincide in the regime of common validity of both --- \ie\ when, on the one hand, the results of \cite{Bhattacharyya:2008mz,Gath:2013qya} are expanded in the limit of large $D$, and on the other hand, the effective theory that we obtain is regarded as a hydrodynamic gradient expansion. 

\item The fluid/gravity theories allow to reconstruct all of the spacetime between the horizon and the boundary of AdS or Minkowski. The large-$D$ effective theory would seem to be more limited since it reconstructs only a region of radial extent $\sim 1/D$ above the horizon. However, the large-$D$ near-horizon dynamics is decoupled to all perturbative orders from the far-zone \cite{Emparan:2014aba}, so the spacetime out to the asymptotic boundary is simply the unperturbed AdS or Minkowski background.

\item Both effective theories are limited to wavenumbers $\bk\ll D/r_0$ (where $r_0$ is the horizon radius of the black brane). For hydrodynamics, this limitation is set by the temperature scale $\bT \sim D/r_0$ of the fluid, while for the large-$D$ expansion we regard it as coming from the size $\mc{O}(D/r_0)$ of radial gradients close to the horizon. But even if neither the fluid/gravity nor the large-$D$ effective theories reach up to momenta $\bk\sim D/r_0$, we identify important phenomena on black branes that appear for $\bk\sim \sqrt{D}/r_0$. At these scales, the large-$D$ expansion includes all powers of $\bk$, while fluid/gravity hydrodynamics can only capture them order by order. This distinction constitutes the main of advantage of the large-$D$ theory as compared to hydrodynamic formulations of black brane dynamics.

\end{itemize}

The usefulness of the large-$D$ effective theory in capturing finite-wavelength phenomena is shown in this article in the following:
\begin{itemize}
\item AF black branes, both neutral and charged, suffer from the Gregory-Laflamme (GL) instability \cite{Gregory:1993vy}, and we can follow their evolution at large $D$ until they settle down in a stable non-uniform configuration \cite{Horowitz:2001cz,Sorkin:2004qq}. The latter can be viewed as an elastically deformed membrane. When the $p$-brane charge is large enough, the instability disappears.

\item Reissner-Nordstrom AdS black branes are stable, but we find a phenomenon of \textit{charged silence}: the frequency of sound modes is purely imaginary at wavelengths below a charge-dependent value, and thus sound propagation is shut off.
\end{itemize}

Both the existence of static non-uniform black branes and the charged silence phenomenon involve finite wavenumbers $\bk\sim \sqrt{D}/r_0$ and cannot be seen in hydrodynamics.

In sec.~\ref{sec:neutral} we introduce the notion of hydro-elastic complementarity in the large-$D$ effective theory of neutral black branes. In sec.~\ref{sec:RNbranes} we derive the effective theory for Reissner-Nordstrom black branes (AF and AdS), \ie branes with point (Maxwell) charges coupled to a Maxwell field, and develop their fluid and elastic properties. Sec.~\ref{sec:polarbranes} describes briefly the introduction of an electric field and its polarization effect on the black brane. Sec.~\ref{sec:branecharge} gives the large-$D$ effective theory of black $p$-branes with $p$-brane charge, coupled to a $p+1$-form gauge potential. We briefly conclude in sec.~\ref{sec:conclusion}. The appendices contain side remarks and some technical steps; in particular, appendix~\ref{app:derivation} gives a detailed derivation of the effective equations that follows and extends the approach of \cite{Emparan:2015hwa}.
\bigskip

\noindent\textit{Notation and conventions}
\begin{itemize}
\item For AF black $p$-branes we define
\beq
n=D-p-3\,,
\eeq
and for AdS black branes
\beq
n=D-1\,.
\eeq
We will often use $n$ as the large parameter instead of $D$. 

\item Many expressions can be given in unified form by introducing a parameter
\beq\label{AFAdS}
\epsilon=\begin{cases}+1
&\mathrm{for~AF}\,,\\ -1 &\mathrm{for~AdS}\,.
\end{cases}
\eeq

\item We will distinguish physical magnitudes from those of the effective theory, which are appropriately rescaled by powers of $n$, by using boldface for the former, \eg\ the energy density $\brho=n\rho$. Appendix~\ref{app:translation} summarizes the relations between them.

\item We use units where 
$16\pi G=\Omega_{n+1}=\mbox{area of the unit }S^{n+1}.$\footnote{The fact that $G\sim \Omega_{n+1}\sim n^{-n/2}\to 0$ at large $n$ can be related to the vanishing of the gravitational field outside the near-horizon region \cite{Emparan:2013moa,Emparan:2013xia}.}
\end{itemize}

\section{Neutral black brane redux: isothermal fluid, elastic membrane}
\label{sec:neutral}

The basic ideas can be most clearly explained by revisiting the effective theory of neutral black branes, either AF or AdS, that was derived in \cite{Emparan:2015gva}.

Let us begin by identifying how to take the large-$D$ limit. This can be inferred from the properties of static, uniform black branes. In Eddington-Finkelstein coordinates, the AF black brane metric is\footnote{We denote by $t$ the ingoing null coordinate. When $n$ is large, dependence on $t$ is the same as dependence on the asymptotic time that measures time in the effective theory.}
\beq
ds^2=2dt\, dr-\lp 1-\frac{r_0^n}{r^n}\rp dt^2+\delta_{ij}d\bsigma^i d\bsigma^j +r^2d\Omega_{n+1}\,,
\eeq
with spatial indices along the brane
\beq
i,j=1,\dots,D-n-3=p\,,
\eeq
and in AdS
\beq
ds^2=2dt\, dr+r^2\lp -\lp 1-\frac{r_0^n}{r^n}\rp dt^2+\delta_{ij}d\bsigma^id\bsigma^j\rp\,,
\eeq
with
\beq
i,j=1,\dots,D-2=n-1\,.
\eeq
In the latter, the cosmological constant is set to $\Lambda=-n(n-1)/2$. As $D\to\infty$ we will keep $p$ finite in AF, while in AdS we will assume that the metric functions depend on only a finite number of coordinates.

The equation of state relating the energy density $\brho=(n+\epsilon)r_0^n$ and pressure $\bP$ of the branes is (using \eqref{AFAdS})
\beq\label{eos}
\bP=-\epsilon\frac{\brho}{n+\epsilon}\,,
\eeq
 so the speed of sound of long-wavelength perturbations is\footnote{Imaginary $c_s$ for AF black branes corresponds to the GL instability \cite{Emparan:2009at,Camps:2010br}.} 
\beq
c_s=\sqrt{\frac{\partial \bP}{\partial \brho}}=\sqrt{\frac{-\epsilon}{n+\epsilon}}\,.
\eeq
Since $c_s$ is small when $n\gg 1$, we expect the large-$n$ dynamics to be non-relativistic.
This dictates the scalings with $n$ required to capture this physics: we rescale
\beq\label{rescsigma}
\bsigma^i = \frac{\sigma^i}{\sqrt{n}}\,,
\eeq
in order to focus on small, $\mc{O}(1/\sqrt{n})$ lengths along the brane, and in addition consider worldvolume velocities $\mc{O}(1/\sqrt{n})$. Thus we take the metric along brane directions to be\footnote{As already mentioned, this scaling is not covered by the analysis of \cite{Bhattacharyya:2015dva,Bhattacharyya:2015fdk}.}
\beq\label{rescmet}
g_{tt}=\mc{O}(1)\,,\qquad g_{ij},\,g_{ti}=\mc{O}(1/n)\,.
\eeq
Ref.~\cite{Emparan:2015gva} sought solutions in Bondi-type gauge for the AF neutral brane
\beq\label{AFneut}
ds^2=-2\left(u_t dt + \frac{u_i}{n} d\sigma^i\right)dr-Adt^2-\frac{2}{n} C_i d\sigma^idt+\frac1{n}G_{ij}d\sigma^id\sigma^j +r^2d\Omega_{n+1}\,,
\eeq
and the AdS black brane
\beq\label{AdSneut}
ds^2=-2\left(u_t dt + \frac{u_i}{n} d\sigma^i\right)dr+r^2\lp -Adt^2-\frac{2}{n} C_i d\sigma^idt+\frac1{n}G_{ij}d\sigma^id\sigma^j\rp\,.
\eeq
A convenient way to proceed is to integrate over the sphere $\Omega_{n+1}$ in the AF case, and over the cyclic brane directions in AdS, and obtain theories of gravity in the reduced finite-dimensional spacetimes with a dilaton field for the size of the compactified spaces (see \eg\ appendix~B of \cite{Emparan:2015hwa}). The reduced AF and AdS theories can be related by analytic continuation in $n$, so only one of them needs to be explicitly solved \cite{Caldarelli:2012hy,Caldarelli:2013aaa}.
One finds that $1/n$ terms in $G_{ij}$ must be included for consistency of the Einstein equations at this order. 

The solution found in \cite{Emparan:2015gva} has $u_t=-1$, $u_i=0$ (by gauge choice), and
\beq
A=1-\frac{\rho(t,\sigma)}{\sR}\,,\qquad C_i=\frac{p_i(t,\sigma)}{\sR}\,,\qquad G_{ij}=\delta_{ij}+\frac1{n}\frac{p_i(t,\sigma)p_j(t,\sigma)}{\rho(t,\sigma)\sR}\,,
\eeq
and finite radial coordinate
\beq
\sR=r^n\,.
\eeq
In the AdS solution $p_i\neq 0$ only along the finite number of non-cyclic brane directions. 
Furthermore, the Einstein equations with a radial index imply that the collective fields $\rho(t,\sigma)$ and $p_i(t,\sigma)$ must satisfy the effective field equations\footnote{We believe these are the conservation equations of the quasilocal stress-energy tensor at $\sR\to\infty$ (with appropriate subtraction). However, extracting this stress tensor is subtle \cite{Emparan:2015hwa}, so we omit it.}
\beqa
\partial_t \rho -\partial_i\partial^i \rho &=&-\partial_i p^i\,,\label{dtrhon}\\
\partial_t p_i  -\partial_j\partial^j p_i &=&\epsilon\partial_i \rho -\partial^j\lp\frac{p_i p_j}{\rho}
\rp\,.\label{dtpn}
\eeqa
Spatial brane indices $i,j$ are raised and lowered with the flat metric $\delta_{ij}$.

\subsection{Isothermal fluid}
Eqs.~\eqref{dtrhon} and \eqref{dtpn} have the form of continuity equations for $\rho$ and $p_i$, with $\int d^p\sigma\, \rho$ and $\int d^p\sigma\, p_i$ being conserved in time. This suggests that we change the variable $p^i$ to $v^i$ as
\beq\label{ptov}
p_i=\rho v_i+\partial_i\rho\,.
\eeq
Then \eqref{dtrhon} becomes the continuity equation for mass,
\beq\label{contmass}
\partial_t \rho+\partial_i\lp \rho v^i\rp=0\,,
\eeq
and \eqref{dtpn} the momentum-stress equation
\beq\label{dtrhovneut}
\partial_t (\rho v^i) +\partial_j \lp \rho v^i v^j+\tau^{ij}\rp
=0\,,
\eeq
with
\beq\label{stressneut}
\tau_{ij}= -\epsilon \rho\,\delta_{ij}  -2\rho \partial_{(i}v_{j)}- \rho\,\partial_j\partial_i \ln \rho\,.
\eeq
These are the equations of a non-relativistic, compressible fluid with mass density $\rho$, velocity $v^i$, and stress tensor $\tau_{ij}$. 
The first two terms in \eqref{stressneut} correspond, respectively, to isothermal-gas pressure
\beq
P=-\epsilon\rho\,,
\eeq
and to shear and bulk viscosities
\beq
\eta=\rho\,,\qquad 
\zeta=\frac{1+\epsilon}{p}\eta
\eeq
(recall that $\delta^i{}_i=p$ is finite for the AF black $p$-brane, but infinite for the AdS brane). Together with the entropy density and temperature,
\beq\label{sTneut}
s=4\pi\rho\,,\qquad T=\frac1{4\pi}\,,
\eeq
which satisfy
\beq
\rho=Ts\,,\qquad d\rho=Tds\,,
\eeq
these properties reproduce the leading large-$n$ results for AF and AdS black branes in the fluid/gravity correspondences of \cite{Bhattacharyya:2008mz,Camps:2010br}. That is, if we take the large-$n$ limit of the stress-energy tensor of the latter, including up to viscosity terms, and scale physical quantities as in appendix~\ref{app:translation}, then we obtain the first two terms in \eqref{stressneut}. Negative $P$ for $\epsilon=+1$ gives rise to the Gregory-Laflamme instability \cite{Camps:2010br,Emparan:2009at}.

The constitutive relation \eqref{stressneut} contains one last term beyond the viscous stress. In fact, since the large-$D$ expansion and the hydrodynamic gradient expansion are different, one might have expected an infinite number of higher-derivative terms in $\tau_{ij}$. Remarkably, the gradient expansion is truncated at a finite order when $D\to\infty$. This implies that an infinite number of higher-order transport coefficients vanish in this limit  \cite{Camps:2010br}. 
In order to match the last term in \eqref{stressneut} to the hydrodynamic second-order coefficients computed in \cite{Bhattacharyya:2008mz} one must focus on the regime where both expansions agree. Hence we must not only take the large-$n$ limit of  \cite{Bhattacharyya:2008mz}, but also regard \eqref{stressneut} as a perturbative gradient expansion, so that the hydrodynamic equations at first-derivative order can be used to rewrite the second-order term, with gradients of $\rho$, in terms of velocity gradients.

In the hydrodynamic interpretation, the last term in $\tau_{ij}$ is associated to creation or dissipation of density inhomogeneities. But we can also interpret it in other ways. Let us write its divergence as
\beq\label{gradtodiv}
\partial_j\lp \rho\,\partial^j\partial_i \ln \rho\rp
=\rho\partial_i\lp \frac{\partial_j\partial^j\rho}{\rho}-\frac{\partial_j\rho\partial^j\rho}{2\rho^2}\rp\,,
\eeq
\ie\ as a term proportional to the gradient of a potential. In this manner we can view this term as yielding an external gravitational force, proportional to the mass density, acting on the fluid. However, the origin of the gravitational potential from second derivatives of $\rho$ is obscure. Since this rewriting will not be possible for dynamical charged black branes, we shall not dwell anymore on it. In appendix~\ref{app:diffusive} we discuss another rewriting of the equations that leads to a simple but unusual identification of the fluid properties. Next we turn to a different interpretation.

\subsection{Hydro-elastic complementarity}\label{subsec:hydroel}

The variable $\rho$ sets the horizon size $\sR=\sR_h=\rho$ in the black brane solutions \eqref{AFneut}, \eqref{AdSneut}, and as such it also determines the radial position of the effective membrane in the background geometry. Namely, the `near-zone' solutions \eqref{AFneut}, \eqref{AdSneut} are matched to either the Minkowski background
\beq\label{minkback}
ds^2=-dt^2+\frac1{n}\delta_{ij}d\sigma^id\sigma^j+r^2 d\Omega_{n+1}+dr^2\,,\eeq
or to the AdS background 
\beq\label{adsback}
ds^2=-r^2\lp -dt^2+\frac1{n}\delta_{ij}d\sigma^id\sigma^j \rp+\frac{dr^2}{r^2}\,
\eeq
at a `membrane surface' $r=r(\sigma)$, of the form
\beq\label{rnrho}
r^n=\rho(\sigma)\,.
\eeq
When $n$ is large, this is
\beq\label{effsurf}
r=1+\frac{\ln\rho(\sigma)}{n}\,,
\eeq
which describes small, $1/n$ deformations of a uniform surface at $r=1$. The trace of its extrinsic curvature, $K$, and the background gravitational potential $\sqrt{-g_{tt}}$ on it are such that  (see app.~\ref{app:meancurv})
\beq\label{gttK}
\sqrt{-g_{tt}}\,K=n+\frac{1+\epsilon}{2}-\lp\epsilon \ln\rho+\frac{\partial_j\partial^j\rho}{\rho}-\frac{\partial_j\rho\partial^j\rho}{2\rho^2}\rp+\mc{O}(1/n)\,.
\eeq
Let us collect the velocity-independent terms in eq.~\eqref{dtrhovneut}, and use \eqref{gradtodiv} and \eqref{gttK} to write it as
\beq\label{diK}
-\rho\partial_i \lp\sqrt{-g_{tt}}\,K\rp=\partial_t (\rho v_i) +\partial^j \lp \rho v_i v_j -2\rho \partial_{(i}v_{j)}\rp\,.
\eeq
In addition, if we take the time derivative of \eqref{gttK} and use the mass continuity equation \eqref{contmass} we obtain
\begin{equation}\label{dtK}
\rho\partial_t \left(\sqrt{-g_{tt}} K\right) = \partial_i \left(\epsilon\rho v^i + \rho \partial^i \frac{\partial_j \left(\rho v^j\right)}{\rho}\right)\,.
\end{equation}

These equations encapsulate the notion of \textit{hydro-elastic complementarity} in the neutral black brane effective theory. Instead of the hydrodynamic equations for the mass and momentum densities of a fluid we can write the elasticity equations \eqref{diK} and \eqref{dtK} for the time and space derivatives of extrinsic curvature of the membrane, in terms of the radial variable $\rho$ and a velocity field along the membrane. 

The right-hand sides of equations \eqref{diK} and \eqref{dtK} still retain a hydrodynamic flavour. This disappears, however, for configurations with $v^i=0$, which must be static and satisfy the Young-Laplace equation 
\beq
\sqrt{-g_{tt}}\,K=\rm{constant}\,.
\eeq 
Stationary, time-independent configurations, can have $v^i\neq 0$ as long as there is no viscous dissipation, \ie\ $\partial_{(a}v_{b)}=0$, or in other words $\partial_t+v^i\partial_i$ is a Killing vector. In this case, using \eqref{contmass} and \eqref{gradtodiv}, we find that eq.~\eqref{dtrhovneut} becomes
\beq\label{soapeqn0}
\partial_i\lp \frac{v^2}{2}+\epsilon \ln\rho+\frac{\partial_j\partial^j\rho}{\rho}-\frac{\partial_j\rho\partial^j\rho}{2\rho^2}\rp=0\,,
\eeq 
which, to leading non-trivial order at large $n$, is equivalent to
\beq\label{soapeqn}
 \sqrt{-g_{tt}\lp 1-\frac{v^2}{n}\rp }\,K=2\kappa\,.
\eeq
with constant $\kappa$. It is straightforward to check that the latter is actually the surface gravity of the horizon in \eqref{AFneut} and \eqref{AdSneut}. In terms of the physical velocity $\bv$ (see app.~\ref{app:translation}) this equation is
\beq\label{soapeq2}
 \sqrt{-g_{tt}\lp 1-\bv^2\rp }\,K=2\kappa\,,
\eeq
with equality holding at $\mc{O}(n)$ and $\mc{O}(1)$. Thus we have recovered the soap-bubble equation \eqref{soapeq}. It can be viewed as the statement that the surface gravity of the stationary horizon is constant along the brane, even when the horizon itself is not uniform but has locally-varying extrinsic curvature, and when the time coordinate on the brane is locally redshifted by Lorentz-boost and gravitational factors, relative to the canonical time of static observers in the background.

\subsection{Static string solutions}\label{subsec:statstring}

For completeness and later reference, we review here the analysis in \cite{Emparan:2015hwa} of static solutions of \eqref{soapeqn0}, with $v=0$, for which $\rho$ depends on only one spatial coordinate $z$. We refer to these as static string configurations.

Define\footnote{This $\mc{P}$ is twice the one in \cite{Emparan:2015hwa}.}
\beq\label{Plnrho}
\mc{P}(z)=\ln\rho(z)\,.
\eeq
From \eqref{effsurf} we see that $\mc{P}$ measures the $\mc{O}(1/n)$ fluctuation of the radius of the membrane. Eq.~\eqref{soapeqn0} (with $v=0$) can be integrated twice to obtain
\beq
\frac12 \mc{P}'{}^2+U(\mc{P})=E\,,
\eeq
with 
\beq\label{UofP}
U(\mc{P})=\epsilon\lp \mc{P}+\rho_0\, e^{-\mc{P}}\rp\,.
\eeq
Here $E$ and $\rho_0$ are integration constants. We can view this as the classical mechanics of a particle (the undamped Toda oscillator), with position $\mc{P}$, time $z$, potential $U$, and energy $E$. When $\rho_0>0$ the potential has an extremum at $e^{\mc{P}}=\rho_0$.

Trajectories of the particle with $\mc{P}'\neq 0$ correspond to non-uniform string profiles. The potential is dominated by the linear term $\epsilon\mc{P}$ for $\mc{P}>0$, and by $\epsilon\rho_0 e^{-\mc{P}}$ for $\mc{P}<0$. Then, when $\epsilon=-1$ there cannot be any non-trivial, bounded trajectories of the particle, and the only solutions correspond to constant $\mc{P}$ at a maximum of $U$ where $e^{\mc{P}}=\rho=\rho_0>0$. These are uniform AdS black branes.

When $\epsilon=+1$ and $\rho_0>0$ the potential has a minimum where $U=1+\ln\rho_0$. The solution that stays at the minimum is the uniform AF black string, but now periodic trajectories also exist (for $E>1+\ln\rho_0$), which give non-uniform black string solutions. Although the equation cannot be integrated exactly, it is easy to obtain analytical approximations and numerical solutions \cite{Emparan:2015hwa}, which match very well the profiles that result at the end of the time evolution of the dynamical equations \eqref{dtrhon}, \eqref{dtpn} \cite{Emparan:2015gva}.

\section{Reissner-Nordstrom black branes}\label{sec:RNbranes}

We study black brane solutions of
\beq\label{chaction}
I=\int d^D x\sqrt{-g}\lp R-\frac14 F_{MN}F^{MN}-2\Lambda\rp \,,
\eeq
with either $\Lambda=0$ or $\Lambda=-n(n-1)/2$. 
Examination of the properties of the static, uniform black brane solutions of the theory\footnote{See \eg\ appendices A and B of \cite{Caldarelli:2010xz}.} at large $n$ shows that, with the scalings in \eqref{rescsigma}, if the charge is to affect the metric at leading order then the gauge potential must be\footnote{Scaling the gauge field as $A_t=\mc{O}(1/\sqrt{n})$ makes it a test field at leading large-$n$ order. The construction of charged black branes by Kaluza-Klein reduction of the neutral branes of sec.~\ref{sec:neutral} results in this theory.}  
\beq
A_t=\mc{O}(1)\,,\qquad A_i=\mc{O}(1/n)\,.
\eeq $A_r$ can be gauge-fixed to zero. Our metric ansatz is the same as in the neutral case, \ie\ \eqref{AFneut} for AF and \eqref{AdSneut} for AdS. 

We can now solve the Einstein equations for this ansatz perturbatively in the $1/n$ expansion. Direct calculation using computer algebra is a quick and efficient method for getting the solution we give below. However, it is also possible, and at times illuminating, to solve the equations step by step extending the systematic approach of \cite{Emparan:2015hwa} as we explain in appendix~\ref{app:derivation}. 

Our solution to the Einstein and Maxwell equations is
\beq
A=1-\frac{\rho(t,\sigma)}{\sR}+\frac{q(t,\sigma)^2}{2\sR^2}\,,\qquad 
C_i=\left(1-\frac{q(t,\sigma)^2}{2\rho(t,\sigma)\sR}\right) \frac{p_i(t,\sigma)}{\sR}\,,
\eeq 
\begin{align}
G_{ij}&=\delta_{ij}+\frac1{n}\left\{\left( 1-\frac{q(t,\sigma)^2 }{2\rho(t,\sigma)\sR}\right)\frac{p_i(t,\sigma) p_j(t,\sigma)}{\rho(t,\sigma)\sR} 
\right. \nonumber \\
& \hphantom{{} = \delta_{ij}+\frac1{n}\,}
\left.
-\ln\lp 1-\frac{\rho_-(t,\sigma)}{\sR}\rp
\left[ (1+\epsilon)\delta_{ij}+  \partial_i \frac{p_j(t,\sigma)}{\rho(t,\sigma)} +\partial_j \frac{p_i(t,\sigma)}{\rho(t,\sigma)}\right] 
 \right\} \,,
\end{align}
\beq
A_t = -\frac{q(t,\sigma)}{\sR}\,,
\eeq
where the collective fields $\rho(t,\sigma)$, $q(t,\sigma)$, $p_i(t,\sigma)$ of the brane must solve the effective field equations
\beq
\partial_t \rho -\partial_i\partial^i \rho =-\partial_i p^i\,,\label{dtrho1}
\eeq
\beq
\partial_t q-\partial_i\partial^i q = -\partial_i\lp \frac{p^i q}{\rho}\rp\,,\label{dtq1}
\eeq
\beq
\partial_t p_i  -\partial_j\partial^j p_i =-\partial_i (\rho_- -\epsilon \rho_+) -\partial^j\left[\frac{p_i p_j}{\rho}
+\rho_- \lp\partial_i\frac{p_j}{\rho}+\partial_j\frac{p_i}{\rho}\rp\right]\,. \label{dtp1}
\eeq
We have defined
\beq
\rho_\pm =\frac{1}{2}\lp \rho\pm \sqrt{\rho^2-2q^2}\rp\,.
\eeq
The horizon is at $\sR=\rho_+$. We could gauge-transform $A_t$ to vanish there.

In principle the extremal limit $\rho= \sqrt{2}q$ is outside the range of validity of the approximations involved in the derivation. However, the fact that the limit $\sqrt{2}q\to\rho$ of the effective equations appears to be a smooth one (in particular in AdS) is suggestive that they may also be applicable at extremality. Nevertheless, a more careful analysis, not done here, is needed to ascertain this point.\footnote{Ref.~\cite{Davison:2013bxa} argues that hydrodynamics applies to Reissner-Nordstrom AdS$_4$ black branes even at $T=0$, which suggests that the effective large-$D$ theory may have a continuous extremal limit.}

\subsection{Dynamical fluid}

Since eq.~\eqref{dtrho1} is the same as the neutral equation \eqref{dtrhon}, we change again to velocity variables \eqref{ptov}. In addition to the mass-continuity equation \eqref{contmass}, eqs.~\eqref{dtq1} and \eqref{dtp1} take the form of charge continuity
\beq
\label{dtq}
\partial_t q+\partial_i j^i=0 \,,
\eeq
with charge current
\beq\label{chj}
j_i=q v_i -\rho\partial_i\lp\frac{q}{\rho}\rp\,,
\eeq
and the momentum conservation equation \eqref{dtrhovneut} 
with stress tensor
\beq\label{chstress}
\tau_{ij}= \lp\rho_- -\epsilon\rho_+ \rp\delta_{ij}  -2\rho_+\partial_{(i}v_{j)}-(\rho_+-\rho_-)\,\partial_j\partial_i \ln\rho \,.
\eeq
This is a non-relativistic, compressible fluid, with pressure
\beq
P=\rho_- -\epsilon\rho_+\,,
\eeq
shear and bulk viscosities
\beq
\eta=\rho_+\,,\qquad 
\zeta=\frac{1+\epsilon}{p}\eta\,,
\eeq
and a conserved particle number (`baryon charge') with density $q$.
The current \eqref{chj} has a diffusive term, proportional to the gradient of $q/\rho$. If $\rho$ is uniform then \eqref{dtq} is the conventional diffusion equation for $q$, with diffusion constant equal to one, which drives the charge density to smoothen uniformly. More generally, when $\rho$ is not uniform the charge density $q$ diffuses to become proportional to $\rho$. By looking at the equations for a test Maxwell field on the large-$D$ black brane, one sees that this charge diffusion on a horizon is essentially the same as in the membrane paradigm \cite{Damour:1978cg}.

As we discussed in sec.~\ref{sec:neutral}, the truncation of the hydrodynamic expansion to a finite order in gradients is a remarkable consequence of the large-$n$ limit. Up to viscous, first-order gradients, the hydrodynamic theory that we have found for Reissner-Nordstrom AF black branes can be recovered as the large-$n$ (non-relativistic) limit of the first-order hydrodynamics derived for them in \cite{Gath:2013qya,DiDato:2015dia}. However, the last term in $\tau_{ij}$ is new. Unlike in neutral branes, it cannot in general be written as a gravitational force, but as we will see next it is crucial for the elastic interpretation.

\subsection{Elastic membrane}

Eq.~\eqref{dtK} for the time derivative of the extrinsic curvature was obtained using only the mass continuity equation \eqref{contmass} and therefore applies also for RN black branes. However, the spatial derivative of the extrinsic curvature does not seem to take, in the general time-dependent case, as simple a form as \eqref{diK}. We will then assume that the system is in a configuration of stationary equilibrium.

Under the evolution equation \eqref{dtq}, the charge density diffuses until an equilibrium state is reached with
\beq
\partial_i\lp\frac{q}{\rho}\rp=0\,,
\eeq
and hence
\beq\label{constrho}
\partial_i\lp\frac{\rho_\pm}{\rho}\rp=0\,.
\eeq
In addition, when the configuration is time independent and with Killing velocity vector, $\partial_{(i}v_{j)}=0$, eq.~\eqref{dtrhovneut} takes the form
\beq
\rho\partial_i \lp\frac{v^2}2\rp+\partial_j\left(\lp\rho_- -\epsilon\rho_+ \rp\delta^j{}_i -(\rho_+-\rho_-)\partial^j\partial_i\ln\rho \right)=0\,.
\eeq

When $\epsilon=+1$ we can use \eqref{gradtodiv}, \eqref{gttK} and \eqref{constrho} to write this equation as
\beq\label{RNstat}
\partial_i\lp\lp\frac{\rho_+-\rho_-}{\rho}-\frac{v^2}{2n}\rp K\rp =0\,.
\eeq
When $v=0$ this is equivalent to
\beq
K=\frac{2\rho_+}{\rho_+-\rho-}\kappa=\rm{constant},
\eeq
where $\kappa$ is the surface gravity of the black brane. 
The equality holds at $\mc{O}(n)$ and $\mc{O}(1)$.

When $v\neq 0$, and in AdS, $K$ is not simply proportional to a redshifted form of the surface gravity of the black brane. Effective theory magnitudes like $K$ are measured at the membrane location in the `overlap zone' at $\sR\to\infty$, while $\kappa$ is measured at the horizon. The relation between the two, as observed in ref.~\cite{Emparan:2015hwa}, is simple for neutral black holes in \eg\ Minkowski backgrounds and/or leading large-$n$ order, but not in general. 

Nevertheless, there still is a neat elastic form of the stationary equation for all RN black branes. For AF black branes, given \eqref{constrho}, we can redefine
\beq\label{rescv}
v=\sqrt{\frac{\rho_+-\rho_-}{\rho}}\;\bar{v}\,,
\eeq 
so that we recover the stationary equation in the form
\beq
\sqrt{1-\frac{\bar{v}^2}{n}}\,K=\rm{constant}\,.
\eeq

When $\epsilon=-1$, instead of \eqref{rescv} we redefine the lengths
\beq\label{adsresc}
\sigma^i=\sqrt{\frac{\rho_+-\rho_-}{\rho}}\;\bar{\sigma}^i\,,
\eeq
and then the stationary equation can be written as
\beq
\sqrt{-g_{tt}\lp 1-\frac{{v}^2}{n}\rp}\,\bar{K}=\mbox{constant}\,,
\eeq
where $\bar{K}$ is the trace of the extrinsic curvature of the embedding in the background metric with brane coordinates $\bar{\sigma}^i$.

Since static charged black branes satisfy the same equation of constant mean curvature as neutral ones, the non-uniform static black string solutions discussed in sec.~\ref{subsec:statstring} give also the profiles for charged black strings.

\subsection{Thermodynamics and the entropy law}\label{subsec:2ndlaw}

For the charged black brane, in addition to the energy density and charge density
\beq
\rho=\rho_+ +\rho_-\,,\qquad q=\sqrt{2\rho_+\rho_-}\,,
\eeq
we have other local variables, namely the entropy density, temperature, and chemical potential,
\beq
s=4\pi\rho_+\,,\qquad T=\frac{\rho_+-\rho_-}{4\pi\rho_+}\,,\qquad \mu=\frac{q}{\rho_+}=\sqrt{\frac{2\rho_-}{\rho_+}}\,.
\eeq
These are obtained in the conventional manner from the black brane solution, and have been renormalized by appropriate factors of $n$ to render them finite in the effective theory (app.~\ref{app:translation}). As such, they satisfy the generic (non-relativistic) thermodynamic equations
\beq\label{chthermo}
\rho=Ts+\mu q\,,\qquad d\rho=Tds+\mu dq\,,
\eeq
as well as black-hole-specific relations such as $\eta/s=1/(4\pi)$.

Using these expressions we can show that
\beq
\partial_i\lp\frac{q}{\rho}\rp=\frac1{4\pi}\lp\frac{Ts}{\rho}\rp^3\partial_i\lp\frac{\mu}{T}\rp
\eeq
and then write the charge current \eqref{chj} in the canonical form 
\beq
j_i=q v_i -\kappa_q\partial_i\lp\frac{\mu}{T}\rp
\eeq
with coefficient
\beq\label{kapp}
\kappa_q=\lp\frac{Ts}{\rho}\rp^2\frac{Ts}{4\pi}\,.
\eeq
A conventional argument now shows that the second law is satisfied in this system. Namely,
the continuity equations for $\rho$ and $q$, \eqref{contmass} and \eqref{dtq}, together with \eqref{chthermo}, imply that
\beq\label{dseq}
\partial_t s+\partial_i\lp s v^i+\kappa_q\frac{\mu}{T}\partial^i\lp\frac{\mu}{T}\rp\rp=\kappa_q\partial_i\lp\frac{\mu}{T}\rp\partial^i\lp\frac{\mu}{T}\rp\,.
\eeq
Then we identify the entropy current 
\beq
j_{(s)}^i=sv^i+\kappa_q\frac{\mu}{T}\partial^i\lp\frac{\mu}{T}\rp\,,
\eeq
and since the right-hand side of \eqref{dseq} is manifestly non-negative, it follows that the total entropy cannot decrease, 
\beq
\partial_t s+\partial_i j_{(s)}^i\geq 0\,.
\eeq
Observe that charge diffusion generates entropy, but viscosity does not: its production of entropy is suppressed by a factor $1/n$. 

When the charge is small, so that $Ts\simeq \rho$, the diffusion coefficient \eqref{kapp} reproduces the value $\kappa_q=\eta/(4\pi)$ associated to the resistivity of neutral black holes in the membrane paradigm \cite{Damour:1978cg}.

\subsection{Quasinormal modes: Gregory-Laflamme instability (AF) and charged silence (AdS)}
Introduce a small perturbation of the static uniform state,
\beqa
\rho&=&\rho_0+\delta\rho\, e^{-i\omega t+ik_j\sigma^j}\,,\\
q&=&q_0+\delta q\, e^{-i\omega t+ik_j\sigma^j}\,,\\
v^i&=&\delta v^i\, e^{-i\omega t+ik_j\sigma^j}\,,
\eeqa
and solve to linear order in the perturbation. Defining the constants 
\beq\label{apm}
a_\pm=\lp\frac{\rho_\pm}{\rho}\rp_0=\frac12 \lp 1\pm\sqrt{1-\frac{2q_0^2}{\rho_0^2}}\rp\leq 1 \,,
\eeq 
we find three different kinds of modes:

\paragraph{Charge diffusion mode.} The perturbation has $\delta q/\delta \rho\neq q_0/\rho_0$ and the frequency is
\beq
\omega=-i k^2\,.
\eeq
This is the expected purely dissipative mode for charge diffusion. 

\paragraph{Shear mode.} The frequency is
\beq
\omega=-i a_+ k^2\,.
\eeq
Since $a_+\leq 1$, we find that charge diffuses more quickly than shear.

\paragraph{Sound modes.} These have $\delta q/\delta \rho= q_0/\rho_0$ (so $q/\rho$ remains constant and there is no charge diffusion) and frequency
\beq
\omega_\pm=\pm i k\sqrt{\epsilon a_+-a_-+k^2 a_-^2}-ia_+ k^2\,.
\eeq
The properties of sound differ markedly depending on whether the brane is AF or AdS:

\subparagraph{AF branes: GL instability.} When $\epsilon=+1$, whenever $k<k_{GL}=1$ the $+$-mode is an unstable mode with real and positive growth rate
\beq
\Omega=-i\omega_+ =k\sqrt{ a_+-a_-+k^2 a_-^2}-a_+ k^2\,.
\eeq
This is the GL instability at large $D$. The threshold wavenumber $k_{GL}$ is the same independently of charge, and for all $0<k<k_{GL}$ the rate $\Omega$ decreases for larger $q_0/\rho_0$. Hence the presence of charge makes the instability \textit{weaker}.

\subparagraph{AdS branes: charged silence.} When $\epsilon=-1$ the sound mode frequencies are
\beq\label{soundrnads}
\omega_\pm=\pm k\sqrt{1-k^2 a_-^2}-ia_+ k^2\,.
\eeq	
At long wavelengths sound propagates with the expected speed $c_s=1$, \ie\ $1/\sqrt{n}$ in physical units. As $k$ increases the effective sound speed is reduced due to viscous damping, until a critical wavenumber is crossed,
\beq\label{critsilence}
k_c=\frac{1}{a_-} \,,
\eeq
beyond which the two sound modes are still stable but their frequency is purely imaginary so sound does not propagate: the brane is silenced at short scales as a consequence of charge. In proper physical scales the critical value is 
\beq\label{physcritsilence}
\bk_c=\frac{\sqrt{n}}{a_-}
\eeq
(in units $r_0=1$), which is a relatively large momentum, but still below the proper scale of momenta $n(a_+-a_-)$ at which the $1/n$ expansion ceases to be applicable.

One might suspect that the effect could be present only when $n\to\infty$, so that at finite $n$ sound at short scales could be attenuated but not completely silenced. However, the phenomenon persists at the next order in $1/n$. We have computed the $1/n$ correction to quasinormal sound frequencies,
\beq
\omega_\pm =\omega^{(0)}_\pm +\frac1{n}\omega^{(1)}_\pm\,,
\eeq
with $\omega^{(0)}_\pm$ the values in \eqref{soundrnads}. We find
\begin{align}
\omega^{(1)}_\pm=&\mp\frac{k}{(a_+-a_-)\sqrt{1-k^2a_-^2}}\biggl[1-k^2-a_-\Bigl( 2-k^2\lp 4-a_+a_-\lp 9-2(a_+-a_-)^2k^2\rp\rp\Bigr)\notag\\
& \hphantom{{} = \mp\frac{k}{(a_+-a_-)\sqrt{1-k^2a_-^2}}}+ k^2(a_++a_-)\ln\frac{a_+}{a_+-a_-}\biggr]\notag\\
& +i\frac{a_+}{a_+-a_-}k^2\left[ 4-9a_-+2(a_+-a_-)^2a_-k^2\right]
\end{align}
(we omit details of the calculation). This correction diverges at $k=k_c$, so we should not trust the result there, but at other momenta it should be reliable as long as $|\omega^{(1)}_\pm|<n$. Under this condition, when $k>k_c$
both frequencies $\omega_\pm$ are purely imaginary.\footnote{Additionally, $\textrm{Re}\,\omega_\pm$ change sign in a narrow interval of $k$ just below $k_c$. However, $n$ may need to be very large for the $1/n$ expansion to be reliable there.} 

So charged silence may be present at finite $n$. However, the possibility must be kept in mind that non-perturbative effects in $1/n$ may change this conclusion. Unfortunately, there are no independent calculations at finite $n$ that could decide this issue. Ref.~\cite{Davison:2011uk} computed numerically the quasinormal sound frequency for non-extremal RN-AdS$_4$, \ie\ $n=3$, and reached a lowest temperature of $\bT=0.0109\bmu$, for  which the critical momentum  \eqref{physcritsilence} is $\bk_c/\bmu=\sqrt{n a_+/2a_-^3}=2.6$.\footnote{With our normalization \eqref{chaction} for the gauge field, $\bmu$ is twice the one in \cite{Davison:2011uk}.} This is much larger than the largest momentum, $\bk/\bmu=0.25$, reached by the quasinormal mode calculation in \cite{Davison:2011uk}, which did not see the effect.

In the effective theory the phenomenon originates in the damping of sound by viscosity, which given the truncation of the hydrodynamic expansion at a low gradient order, is not inhibited by other stronger short-distance physics on the brane.

There is some similarity between this phenomenon and the disappearance of zero-sound at the transition between the collisionless regime and the hydrodynamic regime, first examined holographically in \cite{Karch:2008fa}. However, zero-sound disappears as a result of increasing thermal excitations, while in our case hydrodynamic sound is silenced as charge is added and hence the temperature decreases, so a different microscopic mechanism would seem to be behind it.

We have also found that the same effect is present, also including $1/D$ corrections, for spherical RN-AdS black holes: the lowest gravitational scalar quasinormal modes become purely damped at finite charge for sufficiently large partial-wave number $\ell$.

\subsection{Non-linear evolution}

It is straightforward to solve numerically the effective equations \eqref{dtrho1}, \eqref{dtq1}, \eqref{dtp1}, extending the study in \cite{Emparan:2015gva}. The results are qualitatively the same: thin-enough AF charged black branes are unstable, and they evolve to settle down into static stable non-uniform configurations with constant $K$. Uniform AdS black branes are stable.

\section{Polarized branes}\label{sec:polarbranes}

We now want to consider the effect of an external electric field parallel to the brane. We find that we can include it easily when the gauge field is $A_t=\mc{O}(1/\sqrt{n})$. This implies that the electric charge of the black brane is a test-charge, \ie\ it does not affect the geometry. Nevertheless, the external field does have an effect on the mass, charge and momentum densities on the brane. In particular, the field can polarize the brane, creating a non-trivial distribution of charge even in cases where the total charge vanishes.

With this scaling of the field, we find that the leading order metric takes the same form as in the neutral solution, while the gauge field is
\beq
A_t=\frac1{\sqrt{n}}\lp V(t,\sigma)-\frac{q(t,\sigma)}{\sR}\rp\,,
\eeq
where $V(t,\sigma)$ is the external electric potential. Besides the mass continuity equation \eqref{contmass}, the effective equations are
\beq
\label{dtqE}
\partial_t q+\partial_i j^i=-\partial_i\lp \rho\partial^i V\rp \,,
\eeq
\beq
\label{dtrhovE}
\partial_t (\rho v^i) +\partial_j \lp \rho v^i v^j+ \tau^{ij}\rp=q\partial^i V \,,
\eeq
with the electric current \eqref{chj} and the stress tensor of the neutral black brane \eqref{stressneut}. The electric field $\propto \partial_i V$ exerts a force on the charge on the fluid but it also effectively contributes to the current, polarizing it to equilibrium configurations with non-uniform $q/\rho=V(\sigma)+\rm{const}$.

In a space with $\sigma^i$ periodically identified we can introduce a non-trivial periodic potential $V$, constant in time, and easily solve these equations numerically. For \eg\ a brane initially uniform and with zero total charge, the polarizing effects in AdS show up easily, creating a stable, periodic non-uniform distribution of mass and charge that follow the external field. 

In AF branes there is a competition between the explicit breaking of translational symmetry caused by a non-uniform polarizing field, and the spontaneous breaking due to the GL instability. Numerical study of the evolution under these competing effects shows that when the polarizing field is small, the final stable static black brane is almost unaffected by it, but when the polarizing field is large it can completely dominate the final state. 

In the presence of a polarizing field, static solutions $\rho(z)$ are given by an analysis similar to sec.~\ref{subsec:statstring}. For static configurations we set
\beq 
\frac{q}{\rho}=V(z)-V_0
\eeq
with constant $V_0$. Integrating twice, and assuming periodicity along $z$, we get the equation
\beq
\frac12 \mc{P}'{}^2+U(\mc{P})+e^{-\mc{P}}\int^z d\bar{z}\,e^{\mc{P}(\bar{z})}V'(\bar{z})(V(\bar{z})-V_0) =E\,,
\eeq
with $U$ as in \eqref{UofP}.

\section{Brane-charged black branes}\label{sec:branecharge}

Now we study $p$-brane solutions of the action
\beq
I=\int d^D x\sqrt{-g}\lp R-\frac1{2(p+2)!}H^2_{[p+2]}\rp \,,
\eeq
which carry $p$-brane charge under the $(p+2)$-form field-strength $H_{[p+2]}=dB_{[p+1]}$. We keep $p$ fixed as the dimension $D=n+p+3$ grows large. 

In this case we do not consider AdS branes: since their worldvolume has $p=D-2$ spatial directions, the field-strength $H_{[p+2]}$ would be a top-form, simply amounting to a renormalization of the cosmological constant.\footnote{We might consider lower-form fields $H_{[p+2]}$ in AdS, with $p$ finite as $D\to\infty$. In general these introduce anisotropic worldvolume dynamics, which is beyond the scope of this article. The dynamics could be truncated to the isotropic sector, but we shall not study this either.}

\subsection{Choice of large-$n$ limit}\label{sec:largenp}

In order to get oriented about how to take the large-$n$ limit, we  study first the static uniform solutions. At any finite $n$, refs.~\cite{Caldarelli:2010xz,Emparan:2011hg} give their energy density, pressure, $p$-brane charge, potential, temperature and entropy density, in terms of two parameters $r_0$ and $\alpha$ ($
r_0\geq 0,\, |\alpha|<\infty$), in the form
\beq
\brho=r_0^n\,n\lp 1+N \sinh^2\alpha+\frac1{n}\rp\,,
\eeq
\beq
\bP=-r_0^n(1+n N \sinh^2\alpha)\,,
\eeq
\beq
\bq_p=r_0^n\sqrt{N}n\sinh\alpha\cosh\alpha\,,\qquad
\bPhi_p=\sqrt{N}\tanh\alpha\,,
\eeq
\beq
\bT=\frac{n}{4\pi r_0}(\cosh\alpha)^{-N}\,,\qquad
\bs=4\pi r_0^{n+1} (\cosh\alpha)^{N}\,,
\eeq
where we have defined
\beq
N=\frac{2}{p+1}+\frac2{n}\,.
\eeq
The speed of sound is \cite{Emparan:2011hg}
\beq
c_s^2=\lp \frac{\partial \bP}{\partial \brho}\rp_{\bq_p}=-\frac1{n+1}\frac{1+(2-Nn) \sinh^2\alpha}{1+\lp 2-\frac{Nn}{n+1}\rp\sinh^2\alpha}\,.
\eeq

Let us now take the large-$n$ limit with $\alpha$ fixed, so $\bq_p/\brho$ and $\bP/\brho$ remain $\mc{O}(1)$. Then the speed of sound
\beq
c_s^2=\frac{N \sinh^2\alpha}{1+(2-N)\sinh^2\alpha}+\mc{O}(1/n)
\eeq
is always positive, \ie\ we never expect a GL instability, even though the $p$-brane is in general non-extremal. Moreover, we have $c_s=\mc{O}(1)$ instead of $\mc{O}(1/\sqrt{n})$, so the system is relativistic. While there is nothing wrong with this limit, it is a regime of brane physics different than we are studying in this paper.

A different large-$n$ limit is obtained for small charge $\bq_p/\brho=\mc{O}(1/\sqrt{n})$, \ie\ set
\beq\label{smallch}
\alpha=\frac{\hat\alpha}{\sqrt{n}}
\eeq
and keep $\hat\alpha$ finite. Then
\beq
c_s^2=-\frac{1-N\hat\alpha^2}{n}
\eeq
is non-relativistic at large $n$, and can change sign if the charge is large enough. Therefore, scaling the metric as in \eqref{rescmet} and the gauge potential $B_{[p+1]}$ as 
\beq
B_{t\sigma^1\dots\sigma^p}=\mc{O}\lp n^{-\frac{p+1}{2}}\rp
\eeq
we expect to capture the physics of hydrodynamic sound and the appearance/disappearance of the GL instability.

\subsection{Large $D$ effective theory}

Following these arguments, we take the ansatz \eqref{AFneut} for the metric
and
\beq
B_{t\sigma^1\dots\sigma^p} = n^{-\frac{p+1}{2}} F(t,\sigma,\sR)\,.
\eeq
We find the solution
\beq
A=1-\frac{\rho(t,\sigma)}{\sR}\,,\qquad C_i=\frac{p_i(t,\sigma)}{\sR}\,,\qquad G_{ij}=\delta_{ij}+\frac1{n}\frac{p_i(t,\sigma)p_j(t,\sigma)-q_p^2\delta_{ij}}{\rho(t,\sigma)\sR}\,,
\eeq
\beq
F = \frac{q_p}{\sR}\,,
\eeq
where now $q_p$ is constant and we set $u_t=-1$, $u_i=\mbox{constant}$. We may also gauge-transform $B$ to make it vanish at the horizon $\sR=\rho$. The effective field equations for $\rho$ and $p^i$ are
\beqa
\partial_t \rho -\partial_i\partial^i \rho &=&-\partial_i p^i\,,\label{dtrhopb}\\
\partial_t p_i  -\partial_j\partial^j p_i &=&\partial_i \rho-\partial^j\lp\frac{p_i p_j-q_p^2\delta_{ij}}{\rho}\rp\notag\\
&=&\lp1-\frac{q_p^2}{\rho^2}\rp\partial_i \rho -\partial^j\lp\frac{p_i p_j}{\rho}\rp\,.\label{dtppb}
\eeqa

\subsection{Fluid dynamics}
The change \eqref{ptov} casts these equations into explicitly hydrodynamic form. Besides the usual mass continuity equation \eqref{contmass} from \eqref{dtrhopb}, eq.~\eqref{dtppb} gives 
\beq\label{dtrhovpb}
\partial_t(\rho v_i)+\partial^j\lp -\rho\lp 1+\frac{q_p^2}{\rho^2}\rp\delta_{ij} +\rho v_i v_j-2\rho\partial_{(i} v_{j)}-\rho\partial_i\partial_j\ln\rho\rp=0\,.
\eeq
The only change in the effective fluid relative to the neutral one is in the pressure,
\beq
P=-\rho\lp 1+\frac{q_p^2}{\rho^2}\rp\,.
\eeq
This is indeed the large-$n$ limit of the pressure of the black brane discussed above in sec.~\ref{sec:largenp}, with the appropriate translation between physical and effective magnitudes (app.~\ref{app:translation}). The charge $q_p$ is a global parameter of the fluid only affecting its pressure, and not a local degree of freedom. Nevertheless, we can associate to it a `local potential' \cite{Emparan:2011hg}
\beq
\Phi_p=n^{\frac{p+1}2}\lp B_{t\sigma^1\dots\sigma^p}(\sR\to\infty)-B_{t\sigma^1\dots\sigma^p}(\sR\to\rho)\rp=\frac{q_p}{\rho}\,.
\eeq

Since in the large-$n$ limit as we have taken it, the entropy density and temperature of the effective theory are the same as in the neutral fluid \eqref{sTneut}, $q_p$ and $\Phi_p$ do not enter the first and second law of thermodynamics, and the entropy is conserved.

\subsection{Elastic interpretation}

Eq.~\ref{diK} (with $g_{tt}=-1$) applies again to this system and gives $\partial_t K$, while eq.~\eqref{dtrhovpb} can be written in the hydro-elastic form
\beq
-\rho\partial_i \lp \lp 1-\frac{q_p^2}{2n\rho^2}\rp\,K\rp =\partial_t(\rho v_i)+\partial^j\lp \rho v_i v_j -2\rho\partial_{(i}v_{j)}\rp\,,
\eeq
with the extrinsic curvature of the brane embedding in Minkowski being \eqref{Kmink}.
For stationary branes we can write
\beq\label{stnrypb}
\lp 1-\frac{q_p^2}{2n\rho^2}-\frac{v^2}{2n}\rp\,K=2\kappa\,,
\eeq
which expresses how the surface gravity $\kappa$ remains constant over the entire horizon of the non-uniform, locally boosted brane.

\subsection{Sound mode and GL instability}

The shear mode is the same as in the absence of charge.
Sound modes have frequency
\beq
\omega_\pm=\pm i k\sqrt{1-\frac{q_p^2}{\rho^2}}-ik^2\,.
\eeq
When $q_p<\rho$ the frequency $\omega_+$ presents a GL instability for wavenumber $k$ smaller than
\beq\label{kglpb}
k_{GL}=\sqrt{1-\frac{q_p^2}{\rho^2}}\,,
\eeq
that is, if \eg\ a string has length $L<2\pi/k_{GL}$, in units where the horizon radius is $r_0=1$, then it is linearly stable. 
Note that $q_p=\rho$ is not an extremal limit, which instead corresponds to $\sqrt{N}\bq_p/\brho=1$. Since we are taking $\bq_p/\brho=\mc{O}(1/\sqrt{n})$ we are always far below this limit. That is, the regime we can access of $\rho<q_p\ll \sqrt{n}\rho$ is one of black branes with regular, non-extremal horizons, but stable ones.

Numerical evolution of the non-linear equations \eqref{dtrhopb}, \eqref{dtppb} confirms that small perturbations of uniform black branes with $q_p<\rho$ grow and evolve until a static non-uniform solution is reached, while if $q_p>\rho$ the brane reverts back to the uniform state.

\subsection{Static string solutions}

The analysis of sec.~\ref{subsec:statstring} yields in this case the mechanics of a particle in the potential
\beq
U(\mc{P})=\mc{P}+\frac{\rho_0^2+q_p^2}{\rho_0} e^{-\mc{P}}-\frac{q_p^2}{2}e^{-2\mc{P}}\,,
\eeq
with constant $\rho_0$. Now in order to have a minimum of the potential (where $e^{\mc{P}}=\rho_0$) we need $\rho_0> q_p$, which means that we are in the range where the uniform string is unstable and tends to develop non-uniformities. In this case, the competition between the terms $+e^{-\mc{P}}$ and $-e^{-2\mc{P}}$ makes the oscillations of $\mc{P}$ take a longer time near its smallest values --- that is, the neck of the non-uniform string, where it is thinner, becomes longer as $q_p$ grows.

\section{Concluding remarks}\label{sec:conclusion}

Hydro-elastic complementarity is the manifestation in the limit $D\to\infty$ of what is arguably one of the most basic properties of a black hole, namely, that the same quantity doubles its role as the geometric size and as the mass. In the large-$D$ effective theory of black branes, we can view $\rho(t,\sigma)$ as the local, fluctuating mass density of a fluid, or as the local radius of an elastic soap bubble embedded in the background spacetime. The effective dynamics can then be alternatively regarded as hydrodynamics, like in \eqref{contmass} and \eqref{dtrhovneut}, or as elasticity, like in \eqref{diK}, \eqref{dtK} and \eqref{soapeq2}.

There are several possible extensions of our work, among them the inclusion of a dilaton in the gravitational theory, of several other types of charge on the brane, possible modifications by the presence of hair in AdS, and generalizing the effective theory to solutions of Lovelock gravities \cite{Chen:2015fuf}. 
Perhaps more important for the large-$D$ programme is the extension of these ideas to the effective theory of finite black holes with compact horizons, for which the lowest quasinormal modes are gapped and not hydrodynamic. Stationary black holes are already known to be described by an effective elasticity theory \cite{Emparan:2015hwa,Suzuki:2015iha}, but the complementary view in terms of a simple dynamical theory of continuous mass distributions is absent. It would be interesting if the more general time-dependent effective theory of \cite{Bhattacharyya:2015dva,Bhattacharyya:2015fdk} could be reinterpreted in that manner.

\section*{Acknowledgements}

We are grateful to Simon Gentle, Christiana Pantelidou and Javier Tarr{\'\i}o for conversations. RE and KI are supported by FPA2013-46570-C2-2-P, AGAUR 2009-SGR-168 and CPAN CSD2007-00042 Consolider-Ingenio 2010. KT is supported by JSPS Grant-in-Aid for Scientific Research No.26-3387.

\appendix

\section{Relation between physical and effective-theory magnitudes}\label{app:translation}

In order to obtain finite magnitudes in the effective theory we have scaled by appropriate powers of $n$ the physical magnitudes of the theory, beginning with \eqref{rescsigma}, which indicates that physical length scales along the brane are $1/\sqrt{n}$ times the corresponding lengths in the effective theory. The physical magnitudes discussed in this article, written in boldface, are related to the effective ones as
\begin{align}
\bomega&=\omega\,,&\bk_i&=\sqrt{n}\,k_i\,,&\\
\boldsymbol{\partial}_i&=\sqrt{n}\,\partial_i\,,&\bv^i&=\frac{v^i}{\sqrt{n}}\,,&\\
\brho&= n\rho\,,& \bP&=P\,,&\\
\bT&=nT\,,& \bs&=s\,,&\\
\bq&=nq\,,& \bmu&=\mu\,,&\\
\bq_{p}&=\sqrt{n}\,q_p\,,& \bPhi_{p}&=\frac{\Phi_p}{\sqrt{n}}\,,&\\
\bfeta&=\eta\,,&\bkappa_q&=n\kappa_q\,.
\end{align}
For Maxwell-charged RN black branes we are taking $\bq/\brho=\mc{O}(1)$, whereas for $p$-brane charged branes $\bq_p/\brho=\mc{O}(1/\sqrt{n})$, hence the different scaling of charges and potentials.

Our length units are $r_0=1$, where $r_0$ is measured from the trace of the extrinsic curvature $K$ of the solution at $\sR\to\infty$ so that
\beq
K=\frac{n}{r_0}+\mc{O}(n^0,\sR^{-1})\,.
\eeq

\section{Diffusive velocity frame}\label{app:diffusive}

In dissipative hydrodynamics with a conserved particle number there is an ambiguity in the definition of the velocity field, such that it either follows the energy flow (Landau frame) or the particle flow (Eckart frame). Our definition of $v^i$ in \eqref{ptov} gives \eqref{contmass} and corresponds to the Landau frame. The Eckart velocity frame is defined in terms of the current \eqref{chj} so that $\bar v^i=j^i/q$. Amusingly, for large-$D$ black branes there exists another fairly natural choice of the velocity flow in terms of the variables $p_i$, namely
\beq
\hat{v}_i=\frac{p_i}{\rho}\,.
\eeq
The resulting effective equations are best illustrated for Reissner-Nordstrom (AF or AdS) black branes. Eqs.~\eqref{dtrho1}, \eqref{dtq1}, \eqref{dtp1} take the suggestive form
\beq
\partial_t \rho+\partial_i\lp \rho \hat{v}^i\rp=\partial_i\partial^i \rho\,,
\eeq
\beq
\partial_t q+\partial_i \lp q \hat{v}^i\rp =\partial_i\partial^i q\,,
\eeq
\beq
\partial_t (\rho \hat{v}_i) +\partial^j \lp P \,\delta_{ij} +\rho \hat{v}_i \hat{v}_j +2\rho_-\partial_{(i}\hat{v}_{j)}\rp =\partial_j\partial^j(\rho \hat{v}_i) \,.
\eeq
Now all the equations have got simple diffusive terms in the right-hand side, all with the same diffusion coefficient, while the stress term $\propto \partial_i\partial_j\ln\rho$ has disappeared. 

When $q\neq 0$ there is an apparent negative viscosity $\hat\eta=-\rho_-$ which would seem to yield anti-dissipation, but this effect is offset by the extra diffusive terms. Another peculiarity is that configurations that are static in the Landau-frame are stationary with non-zero velocity $\hat{v}_i=\partial_i\ln\rho$ in the diffusive frame. Whether the existence of this frame for black branes is particularly useful or significant is at present unclear.

\section{Extrinsic curvature of membrane embeddings}\label{app:meancurv}

Let us calculate the area $A$ and volume $V$ of the surface \eqref{effsurf}. In order to ease a bit the notation we use the variable
\beq
\mc{P}(\sigma)=\ln\lp\rho(\sigma)\rp\,.
\eeq
When $n$ is large, in the Minkowski background \eqref{minkback} we have
\beq
A=\frac{\Omega_{n+1}}{n^{p/2}}\int dt\,d^p\sigma\lp1+\frac{\mc{P}(\sigma)}{n}\rp^{n+1}\lp 1+\frac1{2n}(\partial\mc{P})^2\rp\,,
\eeq
\beq
V=\frac{\Omega_{n+1}}{n^{p/2}}\int dt\,d^p\sigma\,\frac1{n+2}\lp1+\frac{\mc{P}(\sigma)}{n}\rp^{n+2}\,,
\eeq
and in the AdS background \eqref{adsback},
\beq
A=n^{(1-n)/2}\int dt\,d^{n-1}\sigma\lp1+\frac{\mc{P}(\sigma)}{n}\rp^{n}\lp 1+\frac1{2n}(\partial\mc{P})^2\rp\,,
\eeq
\beq
V=n^{(1-n)/2}\int dt\,d^{n-1}\sigma\,\frac1{n}\lp1+\frac{\mc{P}(\sigma)}{n}\rp^{n}\,.
\eeq

We compute the trace of the extrinsic curvature $K$ by functional differentiation
\beq
K=\frac{\delta A}{\delta V}=\frac{\delta A}{\delta \mc{P}(\sigma)}\Big/\frac{\delta V}{\delta \mc{P}(\sigma)}\,.
\eeq
In Minkowski we find
\beq\label{Kmink}
K=n+1 -\lp \mc{P}+\partial^2 \mc{P}+\frac12(\partial\mc{P})^2\rp\,,
\eeq
and in AdS
\beq
K=n -\lp \partial^2 \mc{P}+\frac12(\partial\mc{P})^2\rp\,.
\eeq
In AdS there is a non-trivial gravitational redshift on the surface, namely
\beq
\sqrt{-g_{tt}}=1+\frac{\mc{P}(\sigma)}{n}\,.
\eeq
With these results we obtain 
\beq
\sqrt{-g_{tt}}\,K=n+\frac{1+\epsilon}{2}-\lp\epsilon\mc{P}+\partial^2 \mc{P}+\frac12(\partial\mc{P})^2 \rp+\mc{O}(1/n)\,,
\eeq
which is equivalent to \eqref{gttK}.

\section{Systematic derivation of the effective equations}\label{app:derivation}

Here we derive the effective fluid equations (\ref{dtrho1}--\ref{dtp1}) by hand in a systematic way, without recourse to computer algebra. The approach can be regarded as an extension of the derivation that \cite{Emparan:2015hwa} did for the static case.

\subsection{ADM-type formulation}
We adopt an ADM-type formulation, in which 
we decompose the spacetime into the radial direction and the rest,
\begin{eqnarray}
  ds^2 = N^2d{\bar{\rho}}^2 + g_{\mu\nu} \left(dx^\mu + N^\mu d{\bar{\rho}} \right) \left(dx^\nu + N^\nu d{\bar{\rho}} \right).
\end{eqnarray}
In the large $D$ limit the radial gradients are $\ord{D}$, hence,
\begin{eqnarray}
 N \sim \ord{D^{-1}} , \quad N^\mu \sim \ord{D^{-1}}.\label{eq:gauge-assume}
\end{eqnarray}
The Einstein equation is decomposed into the evolution equation,
\begin{eqnarray}
 \fr{N}\partial_{\bar{\rho}} K^\mu{}_\nu +KK^\mu{}_\nu = R^\mu{}_\nu+\delta^\mu{}_\nu \frac{D-1}{L^2} - \fr{N}\nabla^\mu\nabla_\nu N+ \fr{N}\cL_N K^\mu{}_\nu-S^\mu{}_\nu \label{eq:ext-ev},
\end{eqnarray}
and constraint equations,
\begin{eqnarray}
&&K^2-K^\mu{}_\nu K^\nu{}_\mu = R + \frac{(D-1)(D-2)}{L^2}+2\cE,\\
&& \nabla_\mu K^\mu{}_\nu - K_{,\nu} = \cJ_\nu, \label{momconst}
\end{eqnarray}
where $K^\mu{}_\nu$ is the extrinsic curvature on a constant-${\bar{\rho}}$ surface,
\begin{eqnarray}
 K^\mu{}_\nu = \fr{2N}(g^{\mu\lambda}\partial_{\bar{\rho}} g_{\lambda\nu} - \nabla^\mu N_\nu -\nabla_\nu N^\mu). \label{Kdef}
\end{eqnarray}
$\cE$, $\cJ_\mu$ and $S_{\mu\nu}$ are different radial components of the stress tensor,
\begin{eqnarray}
  \cE = T_{\bar\mu\bar\nu}n^{\bar\mu} n^{\bar\nu},\quad
   \cJ_\mu = T_{\bar\alpha \bar\beta} \perp^{\bar\alpha}{}_\mu n^{\bar\beta},\quad
    S^\mu{}_\nu = T_{\bar\alpha\bar\beta}\perp^{\bar\alpha \mu} \perp^{\bar\beta}{}_{\nu}  -\frac{T^{\bar\alpha}{}_{\bar\alpha}}{D-2}\delta^\mu{}_{\nu},
\end{eqnarray}
where $n^{\bar{\mu}} \partial_{\bar\mu} := N^{-1}(\partial_{\bar{\rho}}-N^\mu \partial_\mu)$ is the normal vector and
$\perp^{\bar\mu\bar\nu} :=  \bar{g}^{\bar\mu\bar\nu} - n^{\bar\mu} n^{\bar\nu}$ is the projection tensor.

We also decompose the Maxwell equations.
We define the $(D-1)$-dimensional `electric' and `magnetic' tensors,\footnote{Of course these are not the electric and magnetic fields in the usual sense.}
\begin{eqnarray}
E_\mu = F_{{\bar{\rho}}\mu},\quad B_{\mu\nu} = F_{\mu\nu}.
\end{eqnarray}
For convenience we also define the $D$-field and $H$-field\footnote{In GR-MHD the Maxwell field is decomposed in a similar manner.}
\begin{eqnarray}
&&D^\mu = F^{{\bar{\rho}}\mu} = N^{-2}(E^{\mu}- N^\lambda B_{\lambda}{}^\mu),\label{appdefD}\\
&&H^{\mu\nu} = F^{\mu\nu}=B^{\mu\nu}+N^{-2}N^\nu (E^\mu- N^\lambda B_\lambda{}^\mu)-N^{-2} N^{\mu}(E^\nu-N^\lambda B_\lambda{}^\nu).
\end{eqnarray}
Greek indices are raised and lowered with $g_{\mu\nu}$. The
Maxwell equations are decomposed as
\begin{eqnarray}
&& \fr{N\sqrt{g}} \partial_{\bar{\rho}} (N \sqrt{g} D^\mu) + \fr{N\sqrt{g}}\partial_\nu ( N\sqrt{g}H^{\nu\mu}) = 0,\label{eq:maxwell-evol-D}\\
&& \partial_{\bar{\rho}} B_{\mu\nu} = \partial_\mu E_\nu - \partial_\nu E_\mu,\label{eq:maxwell-evol-B}\\
&& \fr{N\sqrt{g}} \partial_\mu (N\sqrt{g} D^\mu) = 0,\quad \partial_{[\mu} B_{\nu\lambda]}=0.\label{eq:maxwell-const}
\end{eqnarray}
The first two equations give the evolution in the ${\bar{\rho}}$-direction, while the last two are constraints.
One can calculate the energy momentum tensor, $\cE$, $\cJ_\mu$ and $S_{\mu\nu}$, of the Maxwell field,
\begin{eqnarray}
&& \cE = \fr{4}D^\mu E_\mu - \fr{8}H^{\mu\nu} B_{\mu\nu} - \fr{2}N^\mu D^\nu B_{\mu\nu},\\
&& \cJ_\mu = \fr{2}ND^\lambda B_{\mu\lambda},\\
&& S^\mu{}_\nu = \fr{2}(D^\mu E_\nu+H^{\mu\lambda}B_{\nu\lambda}+N^\mu D^\lambda B_{\nu\lambda})-\frac{2D^\lambda E_\lambda + H^{\alpha\beta}B_{\alpha\beta}}{4(D-2)} \delta^\mu{}_\nu.
\end{eqnarray}

\subsection{Solving the equations}

We first solve the equation for the mean curvature, that is the trace of eq.~(\ref{eq:ext-ev}),
\begin{eqnarray}
 \fr{N}\partial_{\bar{\rho}} K +K^2= R+\frac{(D-1)^2}{L^2} - \fr{N}\nabla^2 N+ \fr{N}N^\mu  \partial_\mu K-S^\lambda{}_\lambda.\label{eq:mean}
\end{eqnarray}
The contribution from the Maxwell field is
\begin{eqnarray}
S^\lambda{}_\lambda =- \frac{D^\mu E_\mu}{2(D-2)}+\fr{4}\frac{D-3}{D-2}H^{\mu\nu}B_{\mu\nu}+\fr{2}N^\mu D^\nu B_{\mu\nu}.
\end{eqnarray}
Supposing that the field strength is the order unity, i.e. $E_\mu \sim \ord{1}$ and $B_{\mu\nu}\sim \ord{1}$,
the Maxwell field does not appear in the leading order of eq.~(\ref{eq:mean}).\footnote{If some directions are magnified as $x^\mu \rightarrow \sqrt{D}x^\nu$, then we should assume another scaling, like $E_\mu \sim \ord{D^{-1}}$.}
The situation is similar for the $q$-form field and the massless/massive scalar.\footnote{A large scalar mass as $\ord{D^2}$ will change the situation.}
Therefore, like in the neutral case, we can solve eq.~(\ref{eq:mean}) for the leading order of $K$,
\begin{eqnarray}
 K = \frac{n}{{\sf r}_0(x)}\coth{\bar{\rho}} + \ord{n^0},
\end{eqnarray}
where we set the lapse $N={\sf r}_0(x)/n$ by a choice of gauge.

Now, we introduce an ansatz that represents the deformed black $p$-brane,
\begin{eqnarray}
  g_{\mu\nu}dx^\mu dx^\nu = G_{ab}({\bar{\rho}}, \sigma) d\sigma^a d\sigma^b +\cR_0^2e^{\frac{2\phi({\bar{\rho}},\sigma)}{n+1}}\omega_{IJ}(\varphi) d\varphi^I d\varphi^J, \label{appanzats}
\end{eqnarray}
where $\omega_{IJ}$ is the metric of $S^{n+1}$ sphere and $\cR_0$ is assumed to be constant.
The $p+1$ metric $G_{ab}$ is given by
\begin{eqnarray}
\begin{split}
 G_{ab}d\sigma^a d\sigma^b &= -Adt^2 + \gamma_{ij} (dz^i - u^i dt)(dz^j-u^j dt)\\ 
 &= (-A+u^2) dt^2-2  \tilde u_i dz^i dt + \gamma_{ij} dz^i dz^j\,,
 \end{split}
\end{eqnarray}
where  $ \tilde u_i = \gamma_{ij} u^j$ and $u^2 = \tilde u_i u^i$.
Here, the orders of metric functions are assumed to be 
\begin{eqnarray}
A=\ord{1}, \quad \tilde u_i =\ord{1/n}, \quad \phi=\ord{1} \quad \mbox{and} \quad \gamma_{ij} = \frac{{\sf r}_0^2}{n}\left(\delta_{ij}+\ord{n^{-1}}\right),
\end{eqnarray}
to capture (at a non-linear level) the physics of the lowest quasinormal modes of the black brane --- these are responsible for the hydrodynamic behavior and the Gregory-Laflamme instability.
The shift vector is chosen so that the whole spacetime takes the form of an ingoing-Eddington-Finkelstein metric,
\begin{eqnarray}
 N^t = -\frac{N}{\sqrt{A}},\quad N^i = -\frac{N}{\sqrt{A}}u^i,\quad N^I=0.
\end{eqnarray}
In this gauge, the boundary condition on the apparent horizon is just the regularity of each metric component.

With the metric ansatz~(\ref{appanzats}), the leading order of eq.~(\ref{eq:mean}) becomes
\begin{eqnarray}
 \fr{{\sf r}_0^2} = \fr{\cR_0^2}+\fr{L^2},
\end{eqnarray}
and thus, ${\sf r}_0^2$ is a constant.

Since the other components of Einstein equation have contributions from the Maxwell field at leading order, 
next we must solve part of the leading-order Maxwell equations.
We assume that the field strength has the following order at large $D$,\footnote{One can obtain this scaling by setting the gauge $A_{\bar{\rho}}=0$ and assuming $A_t \sim \ord{1}, A_i \sim \ord{n^{-1}}$.}
\begin{eqnarray}
 E_t \sim \ord{1},\quad E_i \sim \ord{n^{-1}},\quad B_{ti} \sim \ord{1} \quad \mbox{and} \quad B_{ij} \sim \ord{n^{-1}}.\label{eq:max-assume}
\end{eqnarray}
These assumptions lead to 
\begin{eqnarray}
 &&D^t \simeq -A^{-1}N^{-2}E_t \sim \ord{n^2},\nonum
 && D^i \simeq N^{-2}(\gamma^{ij} E_j-A^{-1}u^i E_t+A^{-1/2}N \gamma^{ij} B_{tj})\sim \ord{n^2},
 \label{eq:lo-asum-D} \\
&& H^{ti}  \sim \ord{n} \quad  \mbox{and} \quad H^{ij} \sim \ord{n}.
\end{eqnarray}
The leading order of eq.~(\ref{eq:maxwell-evol-D}) can now be solved as
\begin{eqnarray}
D^t \simeq  \frac{n^2 q (\sigma^a)}{ {\sf r}_0^2 \sqrt{A}  e^\phi }, \quad
D^i \simeq \frac{n^2 q^i(\sigma^a)}{ {\sf r}_0^2 \sqrt{A}  e^\phi }, \label{eq:maxwell-sol-D}
\end{eqnarray}
where the coefficients are chosen for later convenience. 
From the definition of $D^\mu$~(\ref{appdefD}), we have
\begin{eqnarray}
 E_t \simeq - \frac{\sqrt{A}q}{ e^\phi},\quad E_i \simeq \frac{{\sf r}_0}{n \sqrt{A}}\left(\frac{ {\sf r}_0(q_i-qu_i)}{e^\phi}-B_{ti}\right), \label{eq:maxwell-sol-E}
\end{eqnarray}
where $q_i:=\delta_{ij}q^j$ and $\ u_i := \delta_{ij}u^j$.
Substituting eq.~(\ref{eq:maxwell-sol-D}) into the first constraint in eq.~(\ref{eq:maxwell-const}), we can obtain the effective equation for the charge density
\begin{eqnarray}
 \partial_t q + \partial_i q_i = 0 \label{eq:eff-Q-pre}
\end{eqnarray}
where $q_i$ is determined by the regularity of $E_i$, after a part of the leading geometry and $B_{ti}$ are solved.
Then, we can write the contributions of Maxwell field in Einstein equation with $q$ and $q_i$,
\begin{eqnarray}
&& {\sf r}_0^2 S^t{}_t \simeq - \frac{1}{2}n^2 q^2 e^{-2\phi},\quad  {\sf r}_0^2 S^i{}_t \simeq - \frac{1}{2} n^2q_i q e^{-2\phi}, \nonum
&& S^t{}_i \simeq \frac{1}{2} n q A^{-1}(q_i-qu_i) e^{-2\phi}, 
\label{eq:sab-max}\\
&&  {\sf r}_0^2 S^I{}_J \simeq \frac{1}{2} nq^2e^{-2\phi}\delta^I{}_J. \label{eq:sIJ-max}
\end{eqnarray}

With the form in eq.~(\ref{eq:sIJ-max}), we can solve Einstein's equations for the spherical part. 
The $(I,J)$-component of eq.(\ref{eq:ext-ev}) and the definition of extrinsic curvature~(\ref{Kdef}) give
\begin{eqnarray}
  \partial_{\bar{\rho}}^2 \phi + \coth{\bar{\rho}} \partial_{\bar{\rho}} \phi \simeq 1- \frac{1}{2} q^2 e^{-2\phi}\label{eq:phi-q-lo}.
\end{eqnarray}
With the asymptotic condition $\phi \simeq {\bar{\rho}} $ and the regularity at the horizon ${\bar{\rho}}=0$, the solution becomes
\begin{eqnarray}
&& \phi \simeq \ln m + \ln \left[(1-\chi^2)\cosh^2({\bar{\rho}}/2)+\chi^2\right] \label{eq:phi-q-lo-sol} 
\end{eqnarray}
where $\chi$ is an integration function and $q/(\sqrt{2}\chi)(=:m)$  denotes the deformation of the horizon area density.

\subsection{Embedding condition}
We assume ${\sf r}_0=1$ after setting $\cR_0\rightarrow\infty$ for the AF case and $L\rightarrow \infty$ for the AdS case.
Setting $r= e^\frac{\phi}{n+1}$,
we can specify the embedding conditions for both the AF background
\begin{eqnarray}
ds^2 = - dv^2+ 2dvdr+\fr{n}{\sf r}_0^2dz^i dz_i+r^2 d\Omega_{n+1}^2,
\end{eqnarray}
and the AdS background
\begin{eqnarray}
ds^2 = 2dvdr+r^2\left(-dv^2+\fr{n}dz^i dz_i+\delta_{IJ} d\varphi^I d\varphi^J\right).
\end{eqnarray}
Since $\phi$ also depends on $t$ and $z^i$, the linear elements in the embedded frame are mixed
\begin{eqnarray}
 dr = \frac{1}{n} \left[d{\bar{\rho}}+d\ln\left(m-\frac{1}{2}\frac{q^2}{m}\right)\right]+\ord{e^{-{\bar{\rho}}}}.
\end{eqnarray}
However, the second term can be absorbed to the time coordinate by 
\begin{eqnarray}
 dt = dv-\frac{1}{n}d\ln \left[m\left(1-\chi^2\right)\right].
\end{eqnarray}
With this new time coordinate, the boundary conditions become simply
\begin{eqnarray}
 \begin{array}{rc}
 AF: &g_{tt} = -1+ \ord{e^{-{\bar{\rho}}}},\quad u^i = \ord{e^{-{\bar{\rho}}}}\\
AdS:& g_{tt} = -1-\frac{2}{n}\phi+ \ord{e^{-{\bar{\rho}}}},\quad u^i = \ord{e^{-{\bar{\rho}}}}
 \end{array}.\label{asymcondi}
\end{eqnarray}

\subsection{Fluid equations}

Now we can solve $(\mu,t)$ components of eq.(\ref{eq:ext-ev}) for $K^\mu{}_t$
\begin{eqnarray}
&& K^t{}_t \simeq \frac{n}{{\sf r}_0 \sinh{\bar{\rho}}} \frac{1+ \chi^2\tanh^2({\bar{\rho}}/2)}{1-\chi^2\tanh^2({\bar{\rho}}/2)}, \label{Ktt}\\
&&K^i{}_t \simeq \frac{n}{{\sf r}_0 \sinh {\bar{\rho}}} \left[  
\frac{2}{q} (q^i-q {\tilde v}^i) \frac{\chi^2\tanh^2({\bar{\rho}}/2)}{1-\chi^2\tanh^2({\bar{\rho}}/2)}
+{\tilde v}^i \frac{1+\chi^2\tanh^2({\bar{\rho}}/2)}{1-\chi^2\tanh^2({\bar{\rho}}/2)}
\right]. 
\end{eqnarray}
Here, the integration function in $K^t{}_t$ is fixed so that the asymptotic condition~(\ref{asymcondi}) is satisfied later.
Solving $(\mu,t)$ components of eq.(\ref{Kdef}), we have
\begin{eqnarray}
&& A \simeq \frac{ {\sf r}_0^2 \left[1-\chi^2\right]^2\tanh^2({\bar{\rho}}/2)}{\left[1-\chi^2\tanh^2({\bar{\rho}}/2)\right]^2}, \label{A}\\
&& u^i \simeq \frac{\left[1+\chi^4\tanh^2({\bar{\rho}}/2)\right]{\tilde v}^i-2 \chi^4\tanh^2({\bar{\rho}}/2)(q^i/q)}{\cosh^2({\bar{\rho}}/2)\left[1-\chi^2\tanh^2({\bar{\rho}}/2)\right]^2}
 \nonum
 && \qquad\qquad - \frac{4 \chi\tanh^2({\bar{\rho}}/2)\ln\tanh({\bar{\rho}}/2)}{\left[1-\chi^2\tanh^2({\bar{\rho}}/2)\right]^2}\left(\frac{q^i-q {\tilde v}^i}{q}\chi+\partial^i \chi\right),\label{eq:ui-sol}
\end{eqnarray}
where $\partial^i := \delta^{ij} \partial_j$.

Once $A$ and $\phi$ are obtained, we can also solve Eq.~(\ref{eq:maxwell-evol-B}) for $B_{ti}$
\begin{eqnarray}
B_{ti} \simeq -\frac{ {\sf r}_0 \partial_i \chi \left[1+\chi^2 \tanh^2({\bar{\rho}}/2)\right]}{\left[1-\chi^2\tanh^2({\bar{\rho}}/2)\right]^2\cosh^2({\bar{\rho}}/2)}.
\end{eqnarray}
The second equation of~(\ref{eq:maxwell-const}) gives $\partial_t B_{ij} = \partial_j B_{ti}-\partial_i B_{tj} =\ord{n^{-1}}$, which is consistent with the assumption in eq.~(\ref{eq:max-assume}).
Plugging these back into eq.~(\ref{eq:maxwell-sol-E}), we obtain
\begin{eqnarray}
 E_t \simeq - \frac{  \chi \left[1-\chi^2\right]\tanh({\bar{\rho}}/2) }{ \cosh^2({\bar{\rho}}/2)\left[1-\chi^2\tanh^2({\bar{\rho}}/2)\right]^2},
\end{eqnarray}
In the limit ${\bar{\rho}} \to 0$, the singular part of $E_i$ behaves as 
\begin{eqnarray}
E_i \to \frac{{\sf r}_0}{n} \frac{1}{\left[1-\chi^2\right]}
\left\{ \frac{q_i- q {\tilde v}_i}{q} \chi +  \partial_i \chi \right\} 
\frac{1}{\tanh ({\bar{\rho}}/2)}.
\end{eqnarray}
To avoid the singularity in $E_i$ at the horizon, $q_i$ should be
\begin{eqnarray}
q_i = q{\tilde v}_i - q \frac{\partial_i\chi}{\chi}.\label{eq:qi}
\end{eqnarray}
Then, eq.~(\ref{eq:ui-sol}) becomes
\begin{eqnarray}
 u^i \simeq  \frac{ \left[1- \chi^4 \tanh^2({\bar{\rho}}/2)\right] {\tilde v}^i + \chi^2 \tanh^2({\bar{\rho}}/2) \partial^i \chi^2}{\cosh^2({\bar{\rho}}/2)\left[1-\chi^2\tanh^2({\bar{\rho}}/2)\right]^2}  .
\label{eq:ui-sol-b}
\end{eqnarray}
$E_i$ and $K^i{}_t$ can be expressed as
\begin{eqnarray}
&&E_i\simeq\frac{ {\sf r}_0 \chi\tanh({\bar{\rho}}/2)}{n \cosh^2({\bar{\rho}}/2)  \left[1-\chi^2\tanh^2({\bar{\rho}}/2)\right]} \nonumber\\
&&\qquad\qquad\times
\left[
\left\{1-\chi^2\right\}^2 {\tilde v}_i +   \frac{(1-\chi^2)}{\left[1-\chi^2\tanh^2({\bar{\rho}}/2)\right]} \partial_i \chi^2 \right] \\
&&K^i{}_t\simeq \frac{n}{{\sf r}_0 \sinh {\bar{\rho}}} \left[  
-   \frac{\tanh^2({\bar{\rho}}/2)}{1-\chi^2\tanh^2({\bar{\rho}}/2)} \partial^i\chi^2
+ \frac{1+\chi^2\tanh^2({\bar{\rho}}/2)}{1-\chi^2\tanh^2({\bar{\rho}}/2)} {\tilde v}^i
\right] .
\end{eqnarray}
Substituting eq.~(\ref{eq:qi}) into eq.~(\ref{eq:eff-Q-pre}), we obtain one of the effective equations
\begin{eqnarray}
 \partial_t q+\partial_i (q{\tilde v}^i)  = \partial_i (q \partial^i \ln(\chi)).
 \label{appqeq}
 \end{eqnarray}

From $t$-component of eq.~(\ref{momconst}), we can obtain another effective equation for $m(:=q/(\sqrt{2}\chi))$,
\begin{eqnarray}
\partial_t m +\partial_i (m{\tilde v}^i)=
\frac{ 2 m \left[\partial_i \chi\right]\left[\partial^i \chi\right]  }{1-\chi^2}.\label{appmeq}
\end{eqnarray}
Spatial component of eq.~(\ref{momconst}) gives a time-evoution equation for velocity ${\tilde v}_i$, but we have to solve the next order equations. 
Here, for simplicity, we read the equation from the asymptotic behavior of the spatial component of eq.~(\ref{momconst}).\footnote{
Satisfying the asymptotic equation is a necessary condition.
To see the consistency with the momentum constraint, we have to check whether the equation is satisfied to all orders with respect to $e^{{\bar{\rho}}}$. 
We have checked this by using computer algebra.}

Solving eq.~(\ref{eq:ext-ev}), the asymptotic behaviors of $K^t{}_i$ and $K^i{}_j$ is obtained as 
\begin{eqnarray}
&& K^t{}_i =
 \frac{1}{{\sf r}_0 \sinh({\bar{\rho}})} 
\left[
-\frac{1+\chi^2}{1-\chi^2 }{\tilde v}_i + \frac{\partial_i\chi^2}{1-\chi^2} \right] 
+\ord{n^{-1},e^{-2{\bar{\rho}}}} ,\\
&& K^i{}_j =
\frac{1}{{\sf r}_0 }\left[ -\partial^i\partial_j \ln\left[m\left(1-\chi^2\right)\right]+ \frac{1-\epsilon}{2}\delta^i_j\right]  
\nonumber\\
&& \quad
+\frac{1}{{\sf r}_0 \sinh ({\bar{\rho}}) }
\left[
-\frac{1+\chi^2}{1-\chi^2}{\tilde v}^i {\tilde v}_j
+\frac{\partial^i {\tilde v}_j+\partial_j {\tilde v}^i}{1-\chi^2}
+\frac{ {\tilde v}^i \partial_j\chi+{\tilde v}_j \partial^i\chi}{1-\chi^2}\right.\nonumber\\
&& \qquad\qquad\left.
+\frac{\partial^i\chi^2\partial_j\chi^2
}{\left[1-\chi^2\right]^2}
+\partial^i\partial_j\ln\left[m\left(1-\chi^2\right)\right]-\frac{2\chi^2}{1-\chi^2}\delta^i_j -\frac{1-\epsilon}{2}\delta^i_j\right]
\nonumber\\
&&\qquad\qquad\qquad\qquad + \ord{n^{-1},e^{-2{\bar{\rho}}}}. 
\end{eqnarray}
where $\epsilon=2({\sf r}_0^2/\cR_0^2)-1$.

From the asymptotic condition~(\ref{asymcondi}), the next orders of $A$ and $K^t{}_t$ must behave as
\begin{eqnarray}
&&A = {\sf r}_0\left( 1+ \frac{1-\epsilon}{n} \phi \right) + \ord{n^{-2}, e^{-{\bar{\rho}}}}, \\
&&K^t{}_t= \frac{1-\epsilon}{{\sf r}_0} + \ord{n^{-1},e^{-{\bar{\rho}}}}.
\end{eqnarray}
The Maxwell field must fall in the asymptotic region: 
\begin{eqnarray}
&& B_{it} = \ord{e^{-{\bar{\rho}}}}, \quad B_{ij}=\ord{e^{-{\bar{\rho}}}} \quad \mbox{and} \quad D^t=\ord{e^{-{\bar{\rho}}}}.
\end{eqnarray}
Then, we can solve the next order contributions of eq.~(\ref{eq:mean}) for the next order of $K$
\begin{eqnarray}
&&\partial_i K^{(1)}= \frac{-1}{{\sf r}_0} \partial_i 
\left[\partial^2\ln\left[m\left(1-\chi^2\right)\right] 
+ \frac{1}{2}\left(\partial_j \ln\left[m\left(1-\chi^2\right)\right]\right)^2 \right. \nonumber\\
&& \qquad\qquad \left.
+\frac{1+\epsilon}{2}\ln\left[m\left(1-\chi^2\right)\right]
+(1+\epsilon)\frac{2 \chi^2}{1-\chi^2} \frac{1}{\sinh {\bar{\rho}}} \right]
+\ord{e^{-2{\bar{\rho}}}},
\end{eqnarray}
where $K^{(1)}$ is the $\ord{n^0}$-component of $K$.
With these asymptotic behavior, we can obtain the time-evolution equation for ${\tilde v}_i$ from the spatial component of the momentum constraint~(\ref{momconst}):
\begin{eqnarray}
&& \partial_t {\tilde v}_i + {\tilde v}^j \partial_j {\tilde v}_i = \fr{1+\chi^2}\left[\partial^2 {\tilde v}_i+\partial_j \partial_i {\tilde v}^j+(\partial_i {\tilde v}_j+\partial_j {\tilde v}_i)\partial^j \ln m-\left(\partial^j\chi^2\right)(\partial_i {\tilde v}_j-\partial_j {\tilde v}_i)\right]\nonum
 &&\qquad+\frac{1-\chi^2}{1+\chi^2}\partial_i \left[\frac{\partial^2 m}{m}-\frac{(\partial m)^2}{2m^2}+\ln m \right] 
 -\frac{(1-\epsilon)}{1+\chi^2} \partial_i\ln m 
-\partial_i\left[\frac{ 2\left(\partial_j \chi \right)^2}{1-\chi^2}
+\ln\left[1+\chi^2\right]  \right]
 .\nonumber\\
 \label{eq:mom-c-i}
\end{eqnarray}

\subsection{Correspondence with Bondi coordinates}
At leading order we can reproduce the solution in the Bondi coordinates if we employ the following relation among radial coordinates,
\begin{eqnarray}
 {\sf R} = e^\phi  = m \left[\cosh^2({\bar{\rho}}/2)-\chi^2\sinh^2({\bar{\rho}}/2)\right].
\end{eqnarray}
Introducing $\rho$ and $p^i$ as 
\begin{eqnarray} 
 \rho=m \left(1+\chi^2\right),\quad p^i = \left(1+\chi^2\right)(m{\tilde v}^i+\partial^i m).
\end{eqnarray}
eqs.(\ref{appqeq}), (\ref{appmeq}) and (\ref{eq:mom-c-i}) become
\begin{eqnarray}
&& \partial_t \rho-\partial_i \partial^i \rho + \partial_i p^i = 0,\\
&& \partial_t q - \partial_i \partial^i q+\partial_i \left(\frac{q p^i}{\rho}\right)=0,\\
&& \partial_t p_i-\partial^2 p_i = - \partial_i(\rho_--\epsilon \rho_+)-\partial^j \left[\frac{p_ip_j}{\rho}+\rho_-\left(\partial_i\frac{p_j}{\rho}+\partial_j\frac{p_i}{\rho}\right)\right].
\end{eqnarray}
where the horizon positions in ${\sf R}$ are given by $\rho_+=m$ and $\rho_-=m \chi^2$.
These equations are the same as eqs.~(\ref{dtrho1} - \ref{dtp1}).


\end{document}